\documentclass[10pt,onecolumn,aps,prd,preprintnumbers,showpacs,superscriptaddress,nofootinbib,amsmath,amssymb,floats,floatfix,showkeys,notitlepage,longbibliography]{revtex4-2}
\usepackage{orcidlink}
\usepackage{comment}
\usepackage{lipsum}
\usepackage{graphicx}
\usepackage{subfigure}
\usepackage{palatino}
\usepackage{sans}
\usepackage{hyperref}
\hypersetup{colorlinks=true,linkcolor=blue,urlcolor=blue,citecolor=blue}
\usepackage[toc,page]{appendix}
\usepackage[normalem]{ulem}
\usepackage{adjustbox}
\usepackage{latexsym}
\usepackage{amsmath}
\numberwithin{equation}{section}
\usepackage{amssymb}
\usepackage{amsfonts}
\usepackage{dcolumn}
\usepackage{bm}
\usepackage{tikz}
\usetikzlibrary{decorations.pathmorphing}
\usepackage{bigints}
\usepackage{array,tabularx,multirow,booktabs}
\usepackage[tracking=true]{microtype}
\usepackage{soul} %for highlighting
\SetTracking{}{500}
\SetTracking{encoding={*}, shape=sc}{40}
\UseRawInputEncoding %for inputenc error%
\allowdisplaybreaks
\usepackage[utf8]{inputenc}
\usepackage{xcolor} % For colored text if needed

\begin{document} \sloppy

\title{Boundary-only weak deflection angles from isothermal optical geometry}

\author{Ali \"Ovg\"un \orcidlink{0000-0002-9889-342X}}
\email{ali.ovgun@emu.edu.tr}
\affiliation{Physics Department, Eastern Mediterranean University, Famagusta, 99628 North
Cyprus via Mersin 10, Turkiye.}

\author{Reggie C. Pantig \orcidlink{0000-0002-3101-8591}} 
\email{rcpantig@mapua.edu.ph}
\affiliation{Physics Department, School of Foundational Studies and Education, Map\'ua University, 658 Muralla St., Intramuros, Manila 1002, Philippines.}

\begin{abstract}
We develop a boundary only method for computing weak gravitational deflection angles at finite source and receiver distances within the Gauss-Bonnet theorem formulation of optical geometry. Exploiting the fact that the relevant equatorial optical manifold is two dimensional, we introduce isothermal (conformal) coordinates in which the optical metric is locally conformal to a flat reference metric and the Gaussian curvature reduces to a Laplacian of the conformal factor. Such an identity converts the curvature area term in the Gauss-Bonnet theorem into a pure boundary contribution via Green/Stokes-type relations, yielding a deflection formula that depends only on boundary data and controlled closure terms. The residual normalization freedom of the isothermal radius is isolated as an additive freedom in the conformal factor and is shown to leave physical observables invariant, eliminating the need for orbit dependent calibration prescriptions. We explicitly implement the boundary only formalism in weak deflection, where the leading bending reduces to elementary one-dimensional integrals evaluated on a flat reference ray in the conformal plane, with finite distance dependence entering solely through endpoint data. We validate the construction by reproducing finite distance weak deflection for Schwarzschild, deriving the leading finite distance charge correction for Reissner-Nordstr\"om, and applying the same boundary only framework to the Kottler (Schwarzschild-de Sitter) geometry as a representative non-asymptotically flat test case, recovering the standard finite distance expansion including the explicit $\mathcal{O}(\Lambda)$ and mixed $\mathcal{O}(\Lambda M)$ contributions to the total deflection angle.
\end{abstract}


\maketitle

\section{Introduction}\label{sec1}
Gravitational lensing is among the cleanest probes of spacetime geometry: light rays (and, more generally, relativistic massive particles) sample the optical structure of the metric, translating curvature and causal properties into measurable deflection, time delay, magnification, and image morphology. In the strong-field regime near compact objects, the lens map is organized by photon surfaces (photon spheres in spherical symmetry), which control critical curves, relativistic images, and the shadow boundary \cite{Claudel:2000yi,Perlick:2004tq,Cunha:2018acu,Adler:2022qtb}. The classic Schwarzschild analyses established the relativistic image sequence and a concrete dictionary between lensing observables and the underlying spacetime \cite{Virbhadra:1999nm,Bozza:2003cp,Virbhadra:2008ws,Bozza:2010xqn}, while the extension to naked singularities demonstrated that horizonless geometries can generate qualitatively distinct image patterns \cite{Virbhadra:2002ju}. Beyond image positions and magnifications, time delays and magnification centroids furnish additional discriminants for horizon tests and for constraining departures from general relativity \cite{Virbhadra:2007kw}. This motivation has been developed in several directions, including global Gaussian bending constructions and lensing/shadow analyses in alternative-gravity settings such as Weyl gravity and modified-gravity black holes; see, e.g., Refs. \cite{Zhang:2024uex,Kasikci:2018mtg,Kuang:2022ojj,Liu:2024wal,Islam:2020xmy,Izmailov:2019uhy,Zhang:2021ygh,Gao:2021luq}. Recent progress has sharpened the characterization of image morphology itself, including detailed studies of Schwarzschild image distortions \cite{Virbhadra:2022iiy} and conserved structures governing distortion across lensed images \cite{Virbhadra:2024xpk}.

From the computational standpoint, a long-standing goal is to treat lensing observables accurately and efficiently across regimes, from weak deflection (where post-Newtonian expansions suffice) to the strong-deflection limit, where the bending angle develops a universal logarithmic divergence near the critical impact parameter \cite{Zhang:2017vap,Azreg-Ainou:2017obt}. The strong-field expansion introduced in \cite{Bozza:2002zj}, its generalizations \cite{Tsukamoto:2016jzh}, and the broad synthesis in \cite{Bozza:2010xqn} provide a standard framework for extracting the relativistic image sequence and its characteristic coefficients for large classes of asymptotically flat, static and spherically symmetric geometries.

A complementary and conceptually powerful viewpoint for weak lensing is geometric and global. In a modern formulation due to Gibbons and Werner \cite{Gibbons:2008rj}, one projects null propagation onto a two dimensional optical manifold and evaluates the total deflection angle via the Gauss-Bonnet theorem: the bending becomes a topological/geometric statement relating curvature and boundary data on the lensing domain \cite{Allendoerfer_1943,Gibbons:2008rj,Takizawa:2020egm}. This approach is attractive because it ties lensing directly to invariant geometric quantities and cleanly separates \emph{bulk} contributions (curvature area terms) from \emph{boundary} contributions (geodesic curvature and corner angles), with well-defined extensions to finite distance configurations \cite{Ishihara:2016vdc,Arakida:2017hrm}. In stationary spacetimes the relevant optical structure becomes Finslerian, and the Kerr problem can be treated within Randers optical geometry \cite{Werner:2012rc}, grounded in the general correspondence between stationary metrics and Zermelo-Randers-Finsler structures \cite{Gibbons:2008zi}. finite distance and gravitomagnetic corrections have been systematically developed for realistic source/observer locations \cite{Ishihara:2016vdc,Ono:2017pie,Ishihara:2016sfv,Ono:2018ybw}, and careful definitions of the \emph{total} deflection angle clarify how the closing curve and boundary terms must be handled to maintain coordinate invariant observables \cite{Arakida:2017hrm}.

Astrophysical lensing is also rarely an idealized vacuum scattering experiment. Dispersive media (notably plasma) generate frequency-dependent propagation and bending, leading to chromatic lensing signatures \cite{Bisnovatyi-Kogan:2010flt,Atamurotov:2021hoq,Alloqulov:2024xak}. Moreover, lensing phenomenology increasingly incorporates massive messengers, including neutrinos and relativistic massive particles, which require a controlled extension of bending and time-delay observables to timelike trajectories \cite{Pang:2018jpm,Li:2019vhp,Li:2021xhy,Alloqulov:2024sns}. These developments motivate formulations that remain geometric, endpoint-operational, and adaptable to nontrivial asymptotics and environments.

Motivated by this broad program, lensing and shadow diagnostics have been applied extensively to regular black holes, wormholes, black-bounce spacetimes, and modified-gravity solutions \cite{Eiroa:2010wm,Shaikh:2019jfr,Nascimento:2020ime,Guo:2022hjp,Wei:2014dka,Kuang:2022ojj, Hui:2025ane,Lu:2025mcm,Lambiase:2024uzy,Capozziello:2025wwl,DeBianchi:2025bgn,Capozziello:2024ucm,Soares:2024rhp,Soares:2023uup,Soares:2023err,Soares:2025hpy,Pereira:2025fvg}. Rotating charged and axion/Kalb-Ramond-inspired geometries provide especially rich joint phenomenology in deflection and shadow observables \cite{Kumar:2019pjp,Kumar:2020hgm}. Relatedly, recent work emphasizes that null-geodesic structure is intertwined with deeper causal and global properties of nonstandard spacetimes, reinforcing lensing as a diagnostic of the underlying geometry rather than merely a scattering observable \cite{Capozziello:2025wwl}.

Finite distance lensing sharpens both conceptual and practical aspects of these questions. In realistic observations the source and receiver are not at infinity, and the bending angle must be defined operationally as a locally measured angle at the endpoints rather than by asymptotic scattering data. Ishihara \emph{et al.} provided a finite distance definition compatible with the Gauss-Bonnet construction and clarified the requisite corner-angle bookkeeping \cite{Ishihara:2016vdc,Ishihara:2016sfv}. Subsequent refinements and applications have demonstrated the importance of boundary choices and background geometry in stationary settings \cite{Ono:2017pie,Ono:2019hkw} and in media-supported lensing where the optical manifold itself is modified \cite{Crisnejo:2018uyn}. In non-asymptotically flat contexts (notably Kottler/Schwarzschild-de Sitter), one must also distinguish lens-induced bending from background geometry effects; this issue has a long history and remains subtle when translated into finite distance observables \cite{Rindler:2007zz,Kasikci:2018mtg}.

Despite these advantages, the standard Gauss-Bonnet implementation typically retains one technically expensive step: evaluating a curvature area integral over a domain whose boundary includes the (generally unknown) light ray. Even in weak deflection, this bulk term can demand explicit curvature expressions and nontrivial integrations on a moving domain \cite{Crisnejo:2018uyn,Ono:2017pie}. There are partial remedies: for example, reducing double integrals to single integrals through carefully engineered primitives and normalizations \cite{Huang:2022iwl}. However, such reductions can become delicate at finite distance, where the closure is anchored at finite endpoints and asymptotic cancellations are unavailable. Related work has also highlighted simplifications based on orbit structure and reference trajectories in weak lensing \cite{Li:2019qyb,Li:2020wvn}, and recent extensions broaden the Gauss-Bonnet lensing toolkit to bound orbits and more practical coordinate choices \cite{Huang:2022gon,Li:2024ujw,Pantig:2026qcf,Ovgun:2026hwi,Ovgun:2026ind}. However, a systematic boundary only reduction of the curvature area term that is both geometrically transparent and naturally compatible with finite distance observables is still highly desirable.

The purpose of the present work is to provide such a reduction within the finite distance Gauss-Bonnet framework. The key observation is structural: the equatorial optical manifold relevant for axisymmetric lensing is two dimensional, and two dimensional Riemannian metrics admit local isothermal coordinates in which the metric is conformal to the flat metric (a consequence of classical isothermal-coordinate existence results; see, e.g., \cite{Chern_1955,Deturck_1981}). In conformal form, the Gaussian curvature is governed by a Laplacian acting on the conformal factor, so the curvature area integral can be converted directly into a boundary integral via Green/Stokes-type identities. This conversion does not rely on selecting special orbits, auxiliary calibration radii, or problem-specific primitives \cite{Li:2019qyb,Li:2020wvn}; it is a generic consequence of two dimensional conformal geometry.

Our contributions are threefold. First, we construct the isothermal-coordinate representation of the equatorial optical metric for the class of static, axisymmetric lensing problems of interest, and we clarify the normalization freedom of the isothermal radius as an additive freedom in the conformal factor that leaves observables invariant. Second, we derive an explicit boundary only representation of the finite distance bending angle in which the bulk curvature term is replaced by boundary evaluations of the conformal factor and its normal derivative, supplemented by geodesic-curvature and corner-angle terms fixed by the chosen closure \cite{Ishihara:2016vdc,Arakida:2017hrm}. Third, we demonstrate an operational weak-deflection implementation: the leading bending reduces to elementary one-dimensional integrals along a reference curve in the conformal plane, and finite distance dependence enters only through endpoint data and the closing prescription \cite{Ono:2019hkw}.

We validate the formalism against canonical examples and then apply it to a representative non-
asymptotically flat potential. In particular, we reproduce the standard finite distance weak bending
for Schwarzschild and obtain the leading finite distance charge correction for Reissner-Nordstr\"om
within the same boundary only framework. We then consider the Kottler geometry as a tractable
application in which the total finite-distance deflection angle must be defined operationally at
finite endpoints: using the boundary only isothermal reduction, we recover the standard finite
distance expansion for the \emph{total} deflection angle (in the sense of
$\alpha=\Psi_R-\Psi_S+\Phi_{RS}$), including the explicit $\mathcal{O}(\Lambda)$ background contribution
and the mixed $\mathcal{O}(\Lambda M)$ term. If one wishes to isolate a lens-induced contribution
in a non-asymptotically flat setting, one may additionally consider a background-subtracted quantity
defined by subtracting the $M=0$ value at fixed $\Lambda$; we emphasize, however, that this
subtraction defines a distinct observable from the total angle \cite{Rindler:2007zz,Ishihara:2016vdc,
Arakida:2017hrm,Kasikci:2018mtg,Ono:2019hkw}.
Finally, we emphasize the scope of the present paper. Although stationary lensing can be formulated using Randers--Finsler optical geometry and related osculating Riemannian constructions, the boundary-only reduction developed here is established only for the static class, where the equatorial optical geometry is genuinely Riemannian. Accordingly, all derivations and applications in this work are restricted to static spacetimes. A systematic stationary extension would require a separate analysis and is left for future work.

\section{Construction of the Cutoff Manifold} \label{sec2}

In the Gauss-Bonnet theorem approach to gravitational lensing, one rephrases null propagation in a four-dimensional Lorentzian spacetime as a geometric problem on an auxiliary Riemannian optical manifold, so that the total deflection can be related to intrinsic curvature and boundary data. This viewpoint underlies the original weak-deflection formulation of Gibbons and Werner and its finite distance refinements. We now set the conventions and geometric framework needed to construct the optical geometry on the equatorial plane.

\subsection{Conventions and setup} \label{sec2.1}
We work on a smooth four-dimensional spacetime manifold $\mathcal{M}$ equipped with a Lorentzian metric $\bar g_{\mu\nu}$ of signature $(-\,+,+,+)$. Greek indices $\mu,\nu,\ldots$ run over $0,1,2,3$, and we employ coordinates $x^\mu=(t,r,\theta,\phi)$ adapted to stationarity and axial symmetry whenever those symmetries are assumed. We use geometrized units $G=c=1$, so that time and length carry the same dimension and masses are measured in units of length. Unless otherwise stated, we restrict attention to future directed null curves relevant for light propagation.

We denote by $\bar\nabla_\mu$ the Levi-Civita connection compatible with $\bar g_{\mu\nu}$. The associated Christoffel symbols are \cite{Wald:1984rg}
\begin{equation}
\bar\Gamma^{\rho}{}_{\mu\nu}
=\frac{1}{2}\bar g^{\rho\lambda}\left(\partial_\mu \bar g_{\nu\lambda}+\partial_\nu \bar g_{\mu\lambda}-\partial_\lambda \bar g_{\mu\nu}\right).
\label{2.1}
\end{equation}
Our curvature conventions follow the standard Wald sign choice, namely \cite{Wald:1984rg}
\begin{equation}
\bar R^{\rho}{}_{\sigma\mu\nu}
=\partial_\mu \bar\Gamma^{\rho}{}_{\nu\sigma}-\partial_\nu \bar\Gamma^{\rho}{}_{\mu\sigma}
+\bar\Gamma^{\rho}{}_{\mu\lambda}\bar\Gamma^{\lambda}{}_{\nu\sigma}
-\bar\Gamma^{\rho}{}_{\nu\lambda}\bar\Gamma^{\lambda}{}_{\mu\sigma},
\label{2.2}
\end{equation}
with Ricci tensor and scalar curvature defined by
\begin{equation}
\bar R_{\mu\nu}=\bar R^{\rho}{}_{\mu\rho\nu},
\qquad
\bar R=\bar g^{\mu\nu}\bar R_{\mu\nu}.
\label{2.3}
\end{equation}
These conventions fix all signs in subsequent curvature expressions and are the ones commonly used in general-relativistic treatments of optical geometry and Gauss-Bonnet lensing.

When stationarity and axial symmetry are imposed, we assume the existence of two commuting Killing fields $k^\mu=(\partial_t)^\mu$ and $m^\mu=(\partial_\phi)^\mu$, so that metric components are independent of $t$ and $\phi$ in the adapted coordinate system. We further assume reflection symmetry across the equatorial plane, $\theta\mapsto\pi-\theta$, so that the hypersurface $\theta=\pi/2$ is invariant under the symmetry and null geodesics can consistently be restricted to this plane for the class of trajectories considered. On the equatorial plane we will ultimately construct a two dimensional Riemannian optical manifold with coordinates $(r,\phi)$; correspondingly, we reserve early Latin indices $a,b,\ldots$ for tensor components intrinsic to that two dimensional optical geometry once it is introduced, while keeping $\mu,\nu,\ldots$ for spacetime tensors throughout.

Finally, we distinguish notationally between spacetime and optical quantities as follows. Objects built from the spacetime metric carry an overbar (e.g., $\bar g_{\mu\nu}$, $\bar\Gamma^{\rho}{}_{\mu\nu}$, $\bar R^{\rho}{}_{\sigma\mu\nu}$). Objects intrinsic to the optical geometry will be written without an overbar (e.g., $g_{ab}$ and its associated curvature) once the optical construction has been performed in the next subsections. This separation will allow us to translate null propagation in $(\mathcal{M},\bar g_{\mu\nu})$ into a purely Riemannian problem on the equatorial optical manifold without ambiguity in symbols.

\subsection{Metric class and equatorial reduction} \label{sec2.2}
We consider a stationary and axisymmetric spacetime $(\mathcal{M},\bar g_{\mu\nu})$ admitting two commuting Killing fields $k^\mu=(\partial_t)^\mu$ and $m^\mu=(\partial_\phi)^\mu$. In coordinates $x^\mu=(t,r,\theta,\phi)$ adapted to these symmetries, the general circular stationary-axisymmetric line element with no explicit dependence on $t$ and $\phi$ and with vanishing $tr$, $t\theta$, $\phi r$, and $\phi\theta$ cross terms can be written as
\begin{equation}
d\bar s^2
=\bar g_{tt}(r,\theta)\,dt^2+2\bar g_{t\phi}(r,\theta)\,dt\,d\phi
+\bar g_{rr}(r,\theta)\,dr^2+\bar g_{\theta\theta}(r,\theta)\,d\theta^2
+\bar g_{\phi\phi}(r,\theta)\,d\phi^2.
\label{2.4}
\end{equation}
We assume $\bar g_{tt}<0$ in the domain of interest and $\bar g_{rr}>0$, $\bar g_{\theta\theta}>0$, $\bar g_{\phi\phi}>0$ away from coordinate singularities, so that the metric has Lorentzian signature and the spatial directions are spacelike.

In addition to stationarity and axial symmetry, we impose reflection symmetry across the equatorial plane, $\theta\mapsto\pi-\theta$, so that $\theta=\pi/2$ is an invariant hypersurface. Under this assumption, null geodesics that start on the equatorial plane with initial tangent vector tangent to that plane remain confined to it. This restriction is standard in the stationary-axisymmetric Gauss-Bonnet lensing literature, where the equatorial optical geometry is the natural arena for defining a two dimensional domain bounded by the light ray and auxiliary curves \cite{Gibbons:2008rj,Arakida:2017hrm,Ono:2019hkw,Gibbons:2015qja}.

Accordingly, we set
\begin{equation}
\theta=\frac{\pi}{2},\qquad d\theta=0,
\label{2.5}
\end{equation}
and define the equatorial metric components
\begin{equation}
\bar g_{tt}(r)\equiv \bar g_{tt}\!\left(r,\frac{\pi}{2}\right),\quad
\bar g_{t\phi}(r)\equiv \bar g_{t\phi}\!\left(r,\frac{\pi}{2}\right),\quad
\bar g_{rr}(r)\equiv \bar g_{rr}\!\left(r,\frac{\pi}{2}\right),\quad
\bar g_{\phi\phi}(r)\equiv \bar g_{\phi\phi}\!\left(r,\frac{\pi}{2}\right).
\label{2.6}
\end{equation}
The induced line element on the equatorial plane becomes
\begin{equation}
d\bar s^2_{\rm eq}
=\bar g_{tt}(r)\,dt^2+2\bar g_{t\phi}(r)\,dt\,d\phi
+\bar g_{rr}(r)\,dr^2+\bar g_{\phi\phi}(r)\,d\phi^2.
\label{2.7}
\end{equation}

Let $x^\mu(\lambda)$ be a future directed null geodesic with affine parameter $\lambda$ and tangent vector $p^\mu\equiv dx^\mu/d\lambda$. Stationarity and axial symmetry imply two conserved quantities along the geodesic, the energy $E$ and azimuthal angular momentum $L$, defined by
\begin{equation}
E\equiv-\,k_\mu p^\mu=-\bar g_{tt}\,\frac{dt}{d\lambda}-\bar g_{t\phi}\,\frac{d\phi}{d\lambda},
\qquad
L\equiv m_\mu p^\mu=\bar g_{t\phi}\,\frac{dt}{d\lambda}+\bar g_{\phi\phi}\,\frac{d\phi}{d\lambda}.
\label{2.8}
\end{equation}
It is convenient to introduce the combination
\begin{equation}
\Delta(r)\equiv \bar g_{t\phi}(r)^2-\bar g_{tt}(r)\,\bar g_{\phi\phi}(r),
\label{2.9}
\end{equation}
which is strictly positive in regions where $t$ is timelike in the sense relevant to stationary spacetimes and where the $(t,\phi)$ block is nondegenerate. Solving Eq. \eqref{2.8} for the coordinate rates yields
\begin{equation}
\frac{dt}{d\lambda}=\frac{E\,\bar g_{\phi\phi}(r)+L\,\bar g_{t\phi}(r)}{\Delta(r)},
\qquad
\frac{d\phi}{d\lambda}=\frac{-E\,\bar g_{t\phi}(r)-L\,\bar g_{tt}(r)}{\Delta(r)}.
\label{2.10}
\end{equation}

Finally, the null condition $\bar g_{\mu\nu}p^\mu p^\nu=0$ on the equatorial plane gives a first integral for the radial motion. Using Eq. \eqref{2.10} in the equatorial line element \eqref{2.7}, we obtain
\begin{equation}
\bar g_{rr}(r)\left(\frac{dr}{d\lambda}\right)^2
=\frac{E^2\,\bar g_{\phi\phi}(r)+2EL\,\bar g_{t\phi}(r)+L^2\,\bar g_{tt}(r)}{\Delta(r)}.
\label{2.11}
\end{equation}
For future directed null curves in asymptotically flat settings one typically takes $E>0$ and introduces the impact parameter $b\equiv L/E$; however, we will defer any weak-field or asymptotic specialization until the optical geometric construction is in place. The relations above are the only spacetime-level inputs needed to pass to the Riemannian optical description used in Gauss-Bonnet lensing.

\subsection{Optical metric and induced two dimensional optical manifold} \label{sec2.3}
The optical geometry relevant for lensing is obtained by rewriting the null condition on the equatorial plane as a statement about the coordinate time lapse $dt$ along a light ray in terms of the spatial increments $(dr\,d\phi)$. This construction is a standard implementation of the relativistic Fermat principle in stationary spacetimes: among all future directed null curves connecting a source worldline to an observer worldline, the actual light ray makes the arrival time stationary, and this variational problem can be recast as a geodesic problem for a suitable Fermat metric on the orbit space of the timelike Killing field \cite{perlick2003ray}.

Starting from the equatorial line element in Eq. \eqref{2.7}, the null condition $d\bar s^2_{\rm eq}=0$ gives a quadratic equation for $dt$:
\begin{equation}
\bar g_{tt}(r)\,dt^2+2\bar g_{t\phi}(r)\,dt\,d\phi+\bar g_{rr}(r)\,dr^2+\bar g_{\phi\phi}(r)\,d\phi^2=0.
\label{2.12}
\end{equation}
We assume $\bar g_{tt}(r)<0$ and $\Delta(r)>0$, where $\Delta(r)$ was defined in Eq. \eqref{2.9}. Solving Eq. \eqref{2.12} for $dt$ and selecting the future-directed null curves yields
\begin{equation}
dt=\beta_\phi(r)\,d\phi+\sqrt{\gamma_{rr}(r)\,dr^2+\gamma_{\phi\phi}(r)\,d\phi^2},
\label{2.13}
\end{equation}
where we have introduced the equatorial coefficients
\begin{equation}
\beta_\phi(r)\equiv-\frac{\bar g_{t\phi}(r)}{\bar g_{tt}(r)},
\qquad
\gamma_{rr}(r)\equiv\frac{\bar g_{rr}(r)}{-\bar g_{tt}(r)},
\qquad
\gamma_{\phi\phi}(r)\equiv\frac{\Delta(r)}{\bar g_{tt}(r)^2}.
\label{2.14}
\end{equation}
By construction, $\gamma_{rr}>0$ and $\gamma_{\phi\phi}>0$ in the region where $\bar g_{tt}<0$ and $\Delta>0$, so that $\gamma_{ab}$ defines a Riemannian metric on the equatorial orbit space with local coordinates $(r,\phi)$.

Equation \eqref{2.13} exhibits the familiar Randers-type structure: the time functional along a spatial curve $C$ on the equatorial plane can be written as
\begin{equation}
T[C]=\int_C \left(\sqrt{\gamma_{ab}(x)\,dx^a dx^b}+\beta_a(x)\,dx^a\right),
\label{2.15}
\end{equation}
with $x^a=(r,\phi)$ and one-form $\beta=\beta_\phi(r)\,d\phi$. The integrand in Eq. \eqref{2.15} is a Finsler function of Randers type, often called the Fermat metric of the stationary spacetime; stationary points of $T[C]$ (with endpoints fixed on source/observer worldlines) correspond to the spatial projections of null geodesics.

For the important subclass of \emph{static} spacetimes, $\bar g_{t\phi}=0$ and hence $\beta_\phi=0$. In that case Eq. \eqref{2.13} reduces to a purely Riemannian relation,
\begin{equation}
dt=\sqrt{g_{ab}(x)\,dx^a dx^b},
\label{2.16}
\end{equation}
where the two dimensional optical metric $g_{ab}$ on the equatorial optical manifold is identified with the coefficients $\gamma_{ab}$:
\begin{equation}
g_{rr}(r)=\frac{\bar g_{rr}(r)}{-\bar g_{tt}(r)},
\qquad
g_{\phi\phi}(r)=\frac{\bar g_{\phi\phi}(r)}{-\bar g_{tt}(r)},
\qquad
g_{r\phi}(r)=0.
\label{2.17}
\end{equation}
We denote the resulting two dimensional Riemannian manifold by $(\mathcal{M}_{\rm opt},g_{ab})$, with local chart $(r,\phi)$ inherited from the spacetime coordinates restricted to $\theta=\pi/2$. In this static setting, the spatial projection of a null geodesic is a geodesic of $(\mathcal{M}_{\rm opt},g_{ab})$ when parameterized by the optical arclength $\ell$ defined through $d\ell^2=g_{ab}dx^a dx^b$, and Eq. \eqref{2.16} shows that $t$ coincides with $\ell$ up to an additive constant along any given ray.

In the remainder of this paper, we restrict attention to the static subclass \(\bar g_{t\phi}=0\), for which the equatorial optical geometry is the Riemannian manifold \((\mathcal{M}_{\rm opt},g_{ab})\) defined by Eq. \eqref{2.17}. Although stationary spacetimes admit a Randers--Finsler optical description, extending the present boundary-only construction to that setting requires additional ingredients and is not part of the present work. For clarity, in a genuinely stationary spacetime the spatial projection of a null geodesic is an extremal of the Randers functional \eqref{2.15}, not a geodesic of the Riemannian metric \(\gamma_{ab}\) alone; when rewritten using the Levi--Civita connection of \(\gamma_{ab}\), its equation contains an additional term built from the two-form \(d\beta\). Accordingly, the geodesic curvature of the projected photon curve with respect to \(\gamma_{ab}\) is generically nonzero, and Eq. \eqref{2.25} applies only after the static specialization \(\beta=0\).

\subsection{Finite distance deflection angle definitions and Gauss-Bonnet formulation} \label{sec2.4}
We now fix a coordinate invariant definition of the gravitational deflection angle when both the source and the receiver are located at finite distances from the lens, and we express this definition geometrically by applying the Gauss-Bonnet theorem to a two dimensional optical manifold. This finite distance viewpoint is essential when asymptotic flatness at spatial infinity is either not available in practice (e.g., astronomical observations at finite ranges) or not used as a simplifying limit in the computation.

Let $(\mathcal{M}_{\rm opt},g_{ab})$ denote the two dimensional optical manifold on the equatorial plane introduced above, with local coordinates $x^a=(r,\phi)$ and optical line element
\begin{equation}
d\ell^2=g_{ab}(x)\,dx^a dx^b.
\label{2.18}
\end{equation}
Along any smooth curve $C\subset\mathcal{M}_{\rm opt}$, we denote the optical arclength by $\ell$ and write its unit tangent as
\begin{equation}
T^a\equiv \frac{dx^a}{d\ell},
\qquad
g_{ab}T^aT^b=1.
\label{2.19}
\end{equation}
We also introduce the (outward) unit normal $N^a$ along $C$, defined by $g_{ab}N^aN^b=1$ and $g_{ab}T^aN^b=0$, with an orientation chosen consistently over the domain introduced below.

To define the finite distance deflection angle, we follow the geometric prescription in which the bending is measured by comparing the photon direction to the local radial direction at the source and at the receiver, both evaluated in the same optical geometry. Concretely, we consider the light ray projected onto $\mathcal{M}_{\rm opt}$ as a curve $\gamma$ connecting the source point $S$ to the receiver point $R$. At each endpoint we introduce the radial coordinate curve (with $\phi$ fixed) and denote its unit tangent, pointing away from the lens, by $e_r^a$. We then define the local optical angles $\Psi_S$ and $\Psi_R$ by
\begin{equation}
\cos\Psi \equiv g_{ab}T^a e_r^b,
\qquad
0\le \Psi \le \pi,
\label{2.20}
\end{equation}
evaluated at $S$ and $R$ along the light ray. In an equatorial polar-type chart $(r,\phi)$, we define the coordinate separation angle between the endpoints by
\begin{equation}
\Phi_{RS}\equiv \phi_R-\phi_S,
\label{2.21}
\end{equation}
with $\phi_S$ and $\phi_R$ the azimuthal coordinates of $S$ and $R$, respectively. (In parts of the finite-distance lensing literature the separation angle is denoted by $\phi_{RS}$; we use the capital symbol $\Phi_{RS}$ to avoid notational collision with the azimuthal coordinate $\phi$.) The finite distance deflection angle $\alpha$ is then defined as \cite{Ishihara:2016vdc,Ishihara:2016sfv}.
\begin{equation}
\alpha \equiv \Psi_R-\Psi_S+\Phi_{RS}.
\label{2.22}
\end{equation}
Here \(\Phi_{RS}\) is not introduced as an independently observable local angle like \(\Psi_S\) and \(\Psi_R\). Rather, it is the azimuthal separation of the source and receiver in the symmetry-adapted polar chart used to define the outward radial directions entering Eq. \eqref{2.20}. Thus Eq. \eqref{2.22} should be understood as the standard Ishihara-type finite-distance definition, in which the physical deflection is encoded in the combination \(\Psi_R-\Psi_S+\Phi_{RS}\), not in \(\Phi_{RS}\) by itself. Its geometric role in the Gauss--Bonnet construction is made explicit later in Sec. \ref{sec5.3} through the corner-angle bookkeeping. This definition reduces to the usual asymptotic deflection angle in the limit where $S$ and $R$ are taken to infinity in asymptotically flat spacetimes, and it remains meaningful at finite distances because $\Psi_S$ and $\Psi_R$ are defined by inner products in the optical geometry rather than by coordinate-dependent direction comparisons.

We next express \(\alpha\) via the Gauss--Bonnet theorem. Let \(D\subset M_{\rm opt}\) be a simply connected region whose boundary \(\partial D\) is piecewise smooth and consists of four segments: the light-ray segment \(\gamma\) from \(S\) to \(R\), two auxiliary endpoint segments \(C_S\) and \(C_R\), and an auxiliary curve \(C_\Gamma\) closing the contour, chosen so that
\[
\partial D=\gamma\cup C_R\cup C_\Gamma\cup C_S
\]
is positively oriented (counterclockwise with respect to the chosen normal). At this stage, the role of \(C_S\) and \(C_R\) is only to encode the local outward radial directions at \(S\) and \(R\) that enter the angle definition in Eq. \eqref{2.20}; they are therefore introduced as radial in this kinematical sense. They are not yet assumed, in full generality, to be geodesics of the optical metric. In the static, rotationally symmetric optical geometries treated later in this paper, one may choose \(C_S\) and \(C_R\) to be coordinate-radial curves \(\phi=\mathrm{const}\), in which case they are simultaneously optical geodesics and their geodesic curvatures vanish. This coincidence is special to that class of geometries and should not be taken as a generic feature of stationary optical manifolds. The Gauss-Bonnet theorem for such a domain reads \cite{Allendoerfer_1943,Chern_1944,Klingenberg2013,Carmo2016}
\begin{equation}
\iint_D K\,dS+\oint_{\partial D} k_g\,d\ell+\sum_{i}\Theta_i = 2\pi,
\label{2.23}
\end{equation}
where $K$ is the Gaussian curvature of $(\mathcal{M}_{\rm opt},g_{ab})$, $dS$ is its area element, $k_g$ is the geodesic curvature of each smooth boundary segment, and the $\Theta_i$ are the exterior (turning) angles at the corner points of $\partial D$ \cite{Gibbons:2008rj}. The geodesic curvature of a boundary curve $C$ with unit tangent $T^a$ and unit normal $N^b$ is defined by
\begin{equation}
k_g \equiv N_b \nabla_T T^b,
\qquad
\nabla_T T^b \equiv T^a\nabla_a T^b,
\label{2.24}
\end{equation}
with $N_b = g_{ab}N^a$, and also where $\nabla$ is the Levi-Civita connection compatible with $g_{ab}$.

For the static case, the light ray projection is a geodesic of the optical metric, so its geodesic curvature vanishes:
\begin{equation}
k_g(\gamma)=0.
\label{2.25}
\end{equation}

To connect Eq. \eqref{2.23} with the finite-distance deflection angle definition Eq. \eqref{2.22}, we choose a (piecewise smooth) finite-distance lensing domain $D\subset\mathcal{M}_{\rm opt}$ whose boundary is decomposed as
\[
\partial D=\gamma\cup C_R\cup C_\Gamma\cup C_S,
\]
with $\gamma$ the photon curve from $S$ to $R$, $C_S$ and $C_R$ auxiliary curves issuing from $S$ and $R$, and $C_\Gamma$ a closing curve connecting their endpoints. For this standard quadrilateral construction, the Gauss-Bonnet theorem \eqref{2.23} can be rearranged into the explicit identity
\begin{equation}
\alpha
= - \iint_D K\,dS \;-\; \int_{C_\Gamma} k_g\,d\ell
\;-\; \int_{C_R} k_g\,d\ell \;-\; \int_{C_S} k_g\,d\ell \;-\; \int_{\gamma} k_g\,d\ell
\;+\; \Phi_{RS},
\label{2.26}
\end{equation}
where $\Phi_{RS}\equiv\phi_R-\phi_S$ and the corner-angle contribution has been expressed in terms of $\Psi_R$ and $\Psi_S$ so that Eq. \eqref{2.26} is equivalent to Eq. \eqref{2.22}. In the static case, \(k_g(\gamma)=0\) (Eq. \eqref{2.25}). Moreover, choosing \(C_S\) and \(C_R\) as optical geodesics does not conflict with the finite-distance angle definition, provided their tangents at \(S\) and \(R\) coincide with the outward radial directions used in defining \(\Psi_S\) and \(\Psi_R\). The corner-angle terms depend only on these endpoint tangent directions, while the remaining dependence on \(C_S\) and \(C_R\) enters through their geodesic-curvature integrals. In the static, rotationally symmetric cases treated later, the coordinate-radial curves \(\phi=\mathrm{const}\) satisfy both requirements, so \(k_g(C_S)=k_g(C_R)=0\).

Equation \eqref{2.23} is the underlying Gauss--Bonnet relation for piecewise smooth domains on the optical manifold, while Eq. \eqref{2.26} is the corresponding finite-distance identity specialized to the present quadrilateral lensing construction. In the remainder of the paper, Eq. \eqref{2.26} serves as the practical starting point: the subsequent task is to choose \(D\) and \(C_\Gamma\) so that the boundary terms are tractable and, crucially, to rewrite the curvature area contribution \(\iint_D K\,dS\) in a form that can be evaluated purely from boundary data in appropriately chosen coordinates.

\section{Isothermal coordinates for the optical metric} \label{sec3}
The Gauss-Bonnet representation of the finite distance deflection angle involves the Gaussian curvature and the geodesic curvature of boundary curves on the optical manifold. In two dimensions, a particularly effective way to control these quantities is to work in \emph{isothermal} (locally conformal) coordinates, in which the optical metric takes a manifestly conformally flat form and the curvature simplifies to a Laplacian of the conformal factor. The existence of such coordinates is a classical result for Riemannian surfaces, tracing back to Gauss and proved in full generality by Korn and Lichtenstein \cite{Korn_1914}.

\subsection{Conformal flatness of two dimensional Riemannian metrics} \label{sec3.1}

The existence of local isothermal coordinates on a two-dimensional Riemannian manifold is a classical result; see, for example, Refs. \cite{Chern_1955,Korn_1914}. Since our later construction uses only the existence of such coordinates, and not the proof, we record here only the defining local form.

On each simply connected patch of $(\mathcal{M}_{\rm opt},g_{ab})$, there exist local coordinates $(u,v)$ in which the metric is conformal to the Euclidean metric. To preserve the downstream equation numbering used in the remainder of the manuscript, we keep the original label for this defining relation:
\begin{equation}
d\ell^2=g_{ab}\,dx^a dx^b = e^{2\varphi(u,v)}\left(du^2+dv^2\right),
\qquad e^{2\varphi}>0,
\label{3.1}
\end{equation}
and we refer to $(u,v)$ as \emph{isothermal coordinates}. In the axisymmetric optical geometries studied in this paper, these local coordinates can be reorganized into isothermal polar coordinates, which is the starting point of the explicit construction in the next subsection. This is a special local fact about the \emph{two-dimensional optical metric} on \((\mathcal{M}_{\rm opt},g_{ab})\); it is not a statement that the full four-dimensional spacetime metric \(\bar g_{\mu\nu}\) is conformally flat. The utility of isothermal coordinates in the present work rests precisely on this two-dimensional simplification.

\subsection{Isothermal radial coordinate and coordinate ansatz} \label{sec3.2}
In the applications of interest, the equatorial optical geometry is axisymmetric, so the optical metric can be written in a polar-type chart $(r,\phi)$ without $r\phi$ cross terms. Accordingly, we take the two dimensional Riemannian line element on $\mathcal{M}_{\rm opt}$ to be of the form
\begin{equation}
d\ell^2 = g_{rr}(r)\,dr^2 + g_{\phi\phi}(r)\,d\phi^2,
\qquad
g_{rr}(r)>0,\ \ g_{\phi\phi}(r)>0,
\label{3.10}
\end{equation}
where $\phi$ has period $2\pi$, and $r$ is a radial coordinate inherited from the underlying spacetime chart restricted to the equatorial plane. The absence of explicit $\phi$-dependence encodes the axial symmetry, and the positivity conditions ensure that Eq. \eqref{3.10} is Riemannian in the region where the optical construction is valid.

The isothermal form in Eq. \eqref{3.1} is most naturally implemented for an axisymmetric metric by adopting \emph{isothermal polar coordinates}. We therefore seek a new radial coordinate $\rho>0$ and a conformal factor $\varphi$ such that the metric becomes conformal to the flat metric in polar form. Concretely, we impose the ansatz
\begin{equation}
d\ell^2 = e^{2\varphi(\rho)}\left(d\rho^2+\rho^2 d\phi^2\right),
\qquad
e^{2\varphi(\rho)}>0,
\label{3.11}
\end{equation}
where $\rho$ is a function of $r$ only, and the angular coordinate $\phi$ is unchanged. This ansatz is compatible with the isothermal definition \eqref{3.1} because, upon introducing local Cartesian coordinates
\begin{equation}
u=\rho\cos\phi,\qquad v=\rho\sin\phi,
\label{3.12}
\end{equation}
one has $du^2+dv^2=d\rho^2+\rho^2 d\phi^2$, so Eq. \eqref{3.11} is exactly Eq. \eqref{3.1} with $\varphi=\varphi(u,v)$ specialized to the axisymmetric case $\varphi=\varphi(\rho)$. For later use, we summarize the notation distinguishing Euclidean reference-geometry objects from optical-geometry objects:
\begin{center}
\begin{tabular}{ll}
Euclidean reference metric on the isothermal plane: & $ds_E^2 \equiv du^2+dv^2 = d\rho^2+\rho^2 d\phi^2$ \\[2pt]
Optical metric in isothermal form: & $d\ell^2 \equiv e^{2\varphi(u,v)}\,ds_E^2$ \\
Arclength elements: & $ds_E$ (Euclidean), \ $d\ell$ (optical) \\
Unit tangent/normal along a boundary curve: & $(\hat t^i,\hat n^i)$ w.r.t.\ $ds_E^2$; \ $(T^a,N^a)$ w.r.t.\ $d\ell^2$ \\
Index conventions used below: & $i,j,\dots$ for $(u,v)$ (Euclidean); \ $a,b,\dots$ for optical $(\mathcal{M}_{\rm opt},g_{ab})$
\end{tabular}
\end{center}
When needed, Euclidean curvature is denoted $k_E$ (computed in $(u,v)$ with $ds_E$), while $k_g$ denotes the geodesic curvature computed using the optical metric $d\ell^2$.

To determine the relation between $r$ and $\rho$, we equate Eqs. \eqref{3.10} and \eqref{3.11} under the change of variables $r=r(\rho)$. Since $dr=(dr/d\rho)\,d\rho$, matching the $d\rho^2$ and $d\phi^2$ coefficients gives
\begin{equation}
g_{rr}(r)\left(\frac{dr}{d\rho}\right)^2=e^{2\varphi(\rho)},
\qquad
g_{\phi\phi}(r)=e^{2\varphi(\rho)}\,\rho^2.
\label{3.13}
\end{equation}
The second relation in Eq. \eqref{3.13} expresses the conformal factor in terms of $\rho$ and the original metric coefficient:
\begin{equation}
e^{2\varphi(\rho)}=\frac{g_{\phi\phi}(r(\rho))}{\rho^2}.
\label{3.14}
\end{equation}
Substituting Eq. \eqref{3.14} into the first relation in Eq. \eqref{3.13} yields an ordinary differential equation for the isothermal radial map:
\begin{equation}
g_{rr}(r)\left(\frac{dr}{d\rho}\right)^2=\frac{g_{\phi\phi}(r)}{\rho^2}.
\label{3.15}
\end{equation}
Equivalently, choosing the positive branch consistent with increasing radial distance, we may write
\begin{equation}
\frac{d\rho}{\rho}=\sqrt{\frac{g_{rr}(r)}{g_{\phi\phi}(r)}}\,dr.
\label{3.16}
\end{equation}
Equation \eqref{3.16} shows that $\rho$ is determined by a single quadrature once the axisymmetric optical metric coefficients are specified. Integrating, we obtain the general implicit solution
\begin{equation}
\rho(r)=\rho_0\exp\!\left(\int_{r_0}^{r}\sqrt{\frac{g_{rr}(\tilde r)}{g_{\phi\phi}(\tilde r)}}\,d\tilde r\right),
\label{3.17}
\end{equation}
where $r_0$ is an arbitrary reference radius and $\rho_0>0$ is an integration constant.

The freedom in choosing $\rho_0$ has a simple geometric meaning. From Eq. \eqref{3.14}, rescaling $\rho\mapsto c\,\rho$ shifts $\varphi$ by an additive constant, $\varphi\mapsto \varphi-\ln c$, while leaving the physical metric $e^{2\varphi}(d\rho^2+\rho^2 d\phi^2)$ unchanged. This scaling freedom will be useful later when we normalize the conformal factor so that flat-space contributions are cleanly isolated in the weak-field regime, without affecting any invariant observable constructed from the optical geometry.

\subsection{Matching conditions and the ODE for the radial map} \label{sec3.3}
The coordinate ansatz in Eq. \eqref{3.11} does not merely define a convenient reparametrization; it imposes a set of consistency conditions that ensure the new coordinate $\rho$ is a valid radial variable on the optical manifold and that the resulting conformal factor is smooth and single-valued on the domain relevant to lensing.

We begin by recalling that the optical metric in the axisymmetric chart $(r,\phi)$ takes the diagonal form \eqref{3.10}, with $g_{rr}(r)>0$ and $g_{\phi\phi}(r)>0$ in the region of interest. These positivity conditions are necessary for two reasons. First, they guarantee that the square root in Eq. \eqref{3.16} is real, so that $\rho(r)$ defined by quadrature is real-valued. Second, they ensure that the mapping $r\mapsto \rho$ can be made strictly monotone on any interval where $g_{rr}/g_{\phi\phi}$ remains finite and continuous, which is precisely the condition required for $r(\rho)$ to exist as a smooth inverse function.

From the matching relations in Eq. \eqref{3.13}, the radial map is determined by the first-order equation
\begin{equation}
g_{rr}(r)\left(\frac{dr}{d\rho}\right)^2=\frac{g_{\phi\phi}(r)}{\rho^2}.
\label{3.18}
\end{equation}
Equation \eqref{3.18} fixes only the square of \(dr/d\rho\), so the local relation is
\[
\frac{dr}{d\rho}=\pm \frac{1}{\rho}\sqrt{\frac{g_{\phi\phi}(r)}{g_{rr}(r)}}.
\]
On the lensing domain considered in this paper, we choose the branch for which the isothermal radius \(\rho\) increases with the original radial coordinate \(r\). Thus,
\begin{equation}
\frac{dr}{d\rho}=\frac{1}{\rho}\sqrt{\frac{g_{\phi\phi}(r)}{g_{rr}(r)}},
\qquad
\frac{d\rho}{dr}=\rho\,\sqrt{\frac{g_{rr}(r)}{g_{\phi\phi}(r)}}.
\label{3.19}
\end{equation}
This branch choice is valid on any interval where \(g_{rr}(r)>0\), \(g_{\phi\phi}(r)>0\), and \(g_{rr}/g_{\phi\phi}\) remains finite and continuous, so that \(r\mapsto\rho\) is monotone and admits a smooth inverse.

The mapping must also be compatible with the global identification $\phi\sim\phi+2\pi$ and with the geometric meaning of $\rho$ as an areal-type radius for the \emph{conformal} background $d\rho^2+\rho^2 d\phi^2$. This imposes a useful regularity requirement: the function $g_{\phi\phi}(r)$ must not vanish on the domain considered, since Eq. \eqref{3.14} would otherwise force the conformal factor to become singular. In typical lensing applications we restrict the optical manifold to the exterior region where the optical metric is Riemannian and where the photon trajectory does not intersect coordinate singularities; this restriction is consistent with excluding, for instance, horizons where the optical metric degenerates or diverges.

The integration constant in Eq. \eqref{3.17} is fixed by a matching condition appropriate to the physical regime. In asymptotically flat spacetimes in the static case, the optical metric approaches the flat Euclidean metric at large $r$ in the sense that
\begin{equation}
g_{rr}(r)\to 1,
\qquad
g_{\phi\phi}(r)\to r^2,
\qquad
r\to\infty.
\label{3.20}
\end{equation}
Under Eq. \eqref{3.20}, the integrand in Eq. \eqref{3.16} behaves as $\sqrt{g_{rr}/g_{\phi\phi}}\to 1/r$, so Eq. \eqref{3.17} implies $\rho(r)\sim C,r$ at large radius for some constant $C>0$. Choosing $C=1$ fixes the residual scaling freedom $\rho\mapsto c\,\rho$ and yields the asymptotic normalization
\begin{equation}
\rho(r)=r\left(1+o(1)\right),
\qquad
r\to\infty,
\label{3.21}
\end{equation}
which in turn implies $\varphi(\rho)\to 0$ from Eq. \eqref{3.14}. This normalization is particularly convenient for weak-deflection calculations, because it ensures that the conformal factor measures the deviation from flat space rather than absorbing an arbitrary constant offset.

At finite source and receiver distances, one may instead fix the constant by prescribing $\rho(r_\ast)=\rho_\ast$ at a reference radius $r_\ast$ inside the domain but outside any excluded region. This yields a unique solution on any interval where Eq. \eqref{3.19} satisfies the standard existence and uniqueness conditions, namely continuity and local Lipschitz continuity of the right-hand side as a function of $r$. In practice, the monotonicity implied by $d\rho/dr>0$ in Eq. \eqref{3.19} guarantees that the transformation is a diffeomorphism between radial intervals and that the optical manifold can be covered by the isothermal polar chart $(\rho,\phi)$ on the lensing domain $D$ used in the Gauss-Bonnet construction.

With the radial map fixed by Eq. \eqref{3.19} and its normalization, the conformal factor $\varphi(\rho)$ is determined algebraically by Eq. \eqref{3.14}. We will make this reconstruction explicit in the next subsection, where we also record the curvature simplifications that are the primary motivation for introducing isothermal coordinates.

\subsection{Conformal factor reconstruction in isothermal form} \label{sec3.4}
Once the isothermal radial map $\rho(r)$ is fixed by the ODE in Eq. \eqref{3.19} together with a normalization condition such as Eq. \eqref{3.21}, the conformal factor in the isothermal polar form is determined directly from the angular metric component. From the matching relation in Eq. \eqref{3.14}, we have
\begin{equation}
e^{2\varphi(\rho)}=\frac{g_{\phi\phi}(r(\rho))}{\rho^2},
\qquad
\varphi(\rho)=\frac{1}{2}\ln g_{\phi\phi}(r(\rho))-\ln\rho.
\label{3.22}
\end{equation}
It is often convenient to regard $\varphi$ as a function of the original coordinate $r$ and then compose with $r=r(\rho)$ when needed. Defining $\varphi(r)\equiv \varphi(\rho(r))$, Eq. \eqref{3.22} becomes
\begin{equation}
\varphi(r)=\frac{1}{2}\ln g_{\phi\phi}(r)-\ln \rho(r).
\label{3.23}
\end{equation}
Differentiating Eq. \eqref{3.23} with respect to $r$ and using Eq. \eqref{3.19} for $d\rho/dr$ gives a useful first-derivative identity,
\begin{equation}
\frac{d\varphi}{dr}
=\frac{1}{2}\frac{d}{dr}\ln g_{\phi\phi}(r)-\sqrt{\frac{g_{rr}(r)}{g_{\phi\phi}(r)}}.
\label{3.24}
\end{equation}
This expression makes explicit how $\varphi$ measures the departure of the optical metric from the flat polar form: the second term in Eq. \eqref{3.24} is exactly the flat-space contribution when $g_{rr}=1$ and $g_{\phi\phi}=r^2$, while the first term encodes the nontrivial radial dependence of the angular scale factor.

In the isothermal coordinates $(\rho,\phi)$, the optical line element is
\begin{equation}
d\ell^2 = e^{2\varphi(\rho)}\left(d\rho^2+\rho^2 d\phi^2\right),
\label{3.25}
\end{equation}
and the associated area element follows from $\det(g_{ab})=e^{4\varphi}\rho^2$:
\begin{equation}
dS = e^{2\varphi(\rho)}\,\rho\,d\rho\,d\phi.
\label{3.26}
\end{equation}
The normalization freedom discussed after Eq. \eqref{3.17} is transparent in this form. A rescaling $\rho\mapsto c\,\rho$ shifts $\varphi$ by a constant, $\varphi\mapsto \varphi-\ln c$, which leaves the physical metric in Eq. \eqref{3.25} invariant. As a consequence, any quantity that depends only on the intrinsic geometry of $(\mathcal{M}_{\rm opt},g_{ab})$ is insensitive to this additive constant. In particular, the Gaussian curvature depends on derivatives of $\varphi$ rather than on its absolute normalization, which will be central when we convert curvature area integrals into boundary terms.

Finally, the flat-space consistency check is immediate. For a Euclidean optical metric, $g_{rr}(r)=1$ and $g_{\phi\phi}(r)=r^2$, Eq. \eqref{3.19} integrates to $\rho=C,r$, and Eq. \eqref{3.22} yields $\varphi=-\ln C$, a pure constant. Choosing the asymptotic normalization $C=1$ gives $\rho=r$ and $\varphi=0$, so Eq. \eqref{3.25} reduces to the standard flat metric in polar coordinates, as required.

\section{Curvature as a Laplacian and reduction to a boundary term} \label{sec4}
In isothermal coordinates the optical metric is conformal to a flat reference metric, and this special structure collapses the intrinsic curvature to a Laplacian acting on the conformal factor. This is the key technical step that allows the Gauss-Bonnet curvature area term to be rewritten as a boundary contribution by elementary integral identities on the plane.

\subsection{Gaussian curvature for conformal metrics and the Laplacian identity} \label{sec4.1}
We work on the two dimensional optical manifold $(\mathcal{M}_{\rm opt},g_{ab})$ and specialize to an isothermal chart $(u,v)$ in which the metric takes the conformal form \cite{Chern_1955}
\begin{equation}
d\ell^2 = e^{2\varphi(u,v)}\left(du^2+dv^2\right),
\qquad e^{2\varphi}>0,
\label{4.1}
\end{equation}
consistent with the construction in Section \ref{sec3}. In these coordinates the reference (flat) metric is $\delta_{ij}$ on $\mathbb{R}^2$ (with $i,j\in{u,v}$), and the physical metric components are
\begin{equation}
g_{ij}=e^{2\varphi}\delta_{ij},
\qquad
g^{ij}=e^{-2\varphi}\delta^{ij},
\qquad
\sqrt{\det g}=e^{2\varphi}.
\label{4.2}
\end{equation}

The Levi-Civita connection of $g_{ij}$ is determined by the Christoffel symbols
\begin{equation}
\Gamma^{k}{}_{ij}
=\frac{1}{2}g^{k\ell}\left(\partial_i g_{j\ell}+\partial_j g_{i\ell}-\partial_\ell g_{ij}\right),
\label{4.3}
\end{equation}
which, upon inserting Eq. \eqref{4.2}, reduce to the standard conformal expression
\begin{equation}
\Gamma^{k}{}_{ij}
=\delta^{k}{}_{i}\,\partial_j\varphi+\delta^{k}{}_{j}\,\partial_i\varphi-\delta_{ij}\delta^{k\ell}\partial_\ell\varphi.
\label{4.4}
\end{equation}
A direct curvature computation using Eq. \eqref{4.4} gives the scalar curvature
\begin{equation}
R = -2e^{-2\varphi}\left(\partial_u^2\varphi+\partial_v^2\varphi\right),
\label{4.5}
\end{equation}
so that the Gaussian curvature $K=R/2$ is
\begin{equation}
K = -e^{-2\varphi}\left(\partial_u^2\varphi+\partial_v^2\varphi\right).
\label{4.6}
\end{equation}
This relation is the special two dimensional case of the general conformal transformation law for curvature and is equivalent to the familiar Liouville-type curvature equation in conformal coordinates (with the overall sign depending only on the Laplacian convention) \cite{Lytchak_2023}.

It is useful to rewrite Eq. \eqref{4.6} in terms of the Laplace-Beltrami operator of the \emph{physical} metric. For a smooth function $f(u,v)$, the Laplace-Beltrami operator is
\begin{equation}
\Delta f \equiv \nabla^i\nabla_i f
=\frac{1}{\sqrt{\det g}}\partial_i\!\left(\sqrt{\det g}\,g^{ij}\partial_j f\right).
\label{4.7}
\end{equation}
Using Eq. \eqref{4.2}, Eq. \eqref{4.7} collapses to
\begin{equation}
\Delta f = e^{-2\varphi}\left(\partial_u^2 f+\partial_v^2 f\right),
\label{4.8}
\end{equation}
which is the flat Laplacian rescaled by the conformal factor. Combining Eqs. \eqref{4.6} and \eqref{4.8}, we arrive at the Laplacian identity
\begin{equation}
K = -\Delta \varphi.
\label{4.9}
\end{equation}
Thus the conformal form \(d\ell^2=e^{2\varphi}(du^2+dv^2)\) does not imply that the optical geometry is flat: all intrinsic curvature is encoded in derivatives of \(\varphi\), and only the special case \(\varphi=\mathrm{const.}\) gives \(K=0\).

Finally, for the axisymmetric isothermal polar coordinates $(\rho,\phi)$ used in Section \ref{sec3}, the flat Laplacian appearing in Eq. \eqref{4.6} becomes
\begin{equation}
\partial_u^2+\partial_v^2
=\partial_\rho^2+\frac{1}{\rho}\partial_\rho+\frac{1}{\rho^2}\partial_\phi^2,
\label{4.10}
\end{equation}
so that the curvature may also be written as
\begin{equation}
K = -e^{-2\varphi(\rho,\phi)}
\left(\partial_\rho^2\varphi+\frac{1}{\rho}\partial_\rho\varphi+\frac{1}{\rho^2}\partial_\phi^2\varphi\right).
\label{4.11}
\end{equation}
In the common axisymmetric case $\varphi=\varphi(\rho)$, Eq. \eqref{4.11} reduces further to
\begin{equation}
K = -e^{-2\varphi(\rho)}\left(\varphi''(\rho)+\frac{1}{\rho}\varphi'(\rho)\right).
\label{4.12}
\end{equation}
which is the form we will use when converting the Gauss-Bonnet curvature area integral into a boundary term.

\subsection{Area to boundary reduction via Green/Stokes-type identities} \label{sec4.2}
Let $D\subset\mathcal{M}_{\rm opt}$ be an oriented, simply connected domain with piecewise smooth boundary $\partial D$. We assume that $\partial D$ consists of a finite number of smooth segments meeting at corner points. The corner contributions are treated separately in the Gauss-Bonnet relation \eqref{2.23}, so in the present subsection we focus on the conversion of the curvature area integral over the smooth interior of $D$ into a boundary integral along the smooth parts of $\partial D$.

We work in an isothermal chart $(u,v)$ in which the optical metric takes the conformal form \eqref{4.1}. Throughout this subsection we assume that the domain $D$ and its (piecewise smooth) boundary $\partial D$ lie within a single isothermal coordinate patch on which $\varphi(u,v)$ is smooth, so that the divergence theorem applies on each smooth boundary segment. Corner points are treated in the Gauss-Bonnet corner sum (the $\sum_i\Theta_i$ term in Eq. \eqref{2.23}) rather than in the Green/Stokes step. If, in a given application, $D$ cannot be covered by one isothermal patch (e.g.\ due to excised points, horizons, or other coordinate obstructions), one may cover $D$ by finitely many isothermal patches and apply the argument patchwise; the internal patch-boundary contributions cancel upon summation, leaving only the contributions from the physical boundary components of $\partial D$ (together with any additional small boundaries introduced by excision). In these coordinates, the area element is
\begin{equation}
dS=\sqrt{\det g}\,du\,dv=e^{2\varphi(u,v)}\,du\,dv.
\label{4.13}
\end{equation}
Using the Laplace-Beltrami representation \eqref{4.7} and the conformal identities \eqref{4.2}, we note a key simplification:
\begin{equation}
\sqrt{\det g}\,g^{ij}=e^{2\varphi}\cdot e^{-2\varphi}\delta^{ij}=\delta^{ij}.
\label{4.14}
\end{equation}
Therefore, for any smooth function $f$,
\begin{equation}
\Delta f\,dS
=\partial_i\!\left(\sqrt{\det g}\,g^{ij}\partial_j f\right)\,du\,dv
=\partial_i\!\left(\delta^{ij}\partial_j f\right)\,du\,dv
=\left(\partial_u^2 f+\partial_v^2 f\right)\,du\,dv.
\label{4.15}
\end{equation}
In particular, combining Eq. \eqref{4.9} with Eq. \eqref{4.15} gives
\begin{equation}
K\,dS = -\Delta\varphi\,dS
= -\left(\partial_u^2\varphi+\partial_v^2\varphi\right)\,du\,dv.
\label{4.16}
\end{equation}
Thus the curvature area density becomes the flat divergence of the flat gradient of $\varphi$ in the $(u,v)$ plane.

Let $\partial D$ be parameterized by a smooth parameter $s_E$ with respect to the \emph{Euclidean} reference metric $du^2+dv^2$, and let $\hat n^i$ denote the outward unit normal to $\partial D$ in this Euclidean sense. The standard Green theorem (divergence theorem in the plane) then yields
\begin{equation}
\iint_D \left(\partial_u^2\varphi+\partial_v^2\varphi\right)\,du\,dv
=\oint_{\partial D} \hat n^i\,\partial_i\varphi\,ds_E.
\label{4.17}
\end{equation}
Substituting Eq. \eqref{4.17} into Eq. \eqref{4.16}, we obtain the desired area-to-boundary reduction:
\begin{equation}
\iint_D K\,dS
= -\oint_{\partial D}\hat n^i\,\partial_i\varphi\,ds_E.
\label{4.18}
\end{equation}

It is useful to rewrite Eq. \eqref{4.18} in intrinsic geometric terms, involving the unit normal and line element associated with the \emph{physical} optical metric. Along the smooth parts of $\partial D$, let $T^a$ be the unit tangent with respect to $g_{ab}$, and $N^a$ the outward unit normal with respect to $g_{ab}$, so that $g_{ab}T^aT^b=1$, $g_{ab}N^aN^b=1$, and $g_{ab}T^aN^b=0$. In the conformal metric \eqref{4.1}, the physical arclength element $d\ell$ and the Euclidean arclength element $ds_E$ are related by
\begin{equation}
d\ell=e^{\varphi}\,ds_E,
\label{4.19}
\end{equation}
and the physical unit normal is related to the Euclidean unit normal by the inverse scaling
\begin{equation}
N^i=e^{-\varphi}\hat n^i.
\label{4.20}
\end{equation}
Using Eqs. \eqref{4.19} and \eqref{4.20}, we find that the Euclidean normal derivative in Eq. \eqref{4.18} equals the intrinsic normal derivative weighted by the physical line element:
\begin{equation}
\hat n^i\,\partial_i\varphi\,ds_E
=\left(e^{\varphi}N^i\partial_i\varphi\right)\left(e^{-\varphi}d\ell\right)
= N^a\nabla_a\varphi\,d\ell,
\label{4.21}
\end{equation}
where we used that $\nabla_a\varphi=\partial_a\varphi$ for a scalar. Substituting Eq. \eqref{4.21} into Eq. \eqref{4.18} gives an invariant form of the reduction:
\begin{equation}
\iint_D K\,dS
= -\oint_{\partial D} N^a\nabla_a\varphi\,d\ell.
\label{4.22}
\end{equation}
Equation \eqref{4.22} is the fundamental identity we require: in isothermal coordinates, the curvature area term appearing in Gauss-Bonnet is converted into a boundary integral of the normal derivative of the conformal factor.

For later use, we also record the specialization to axisymmetric isothermal polar coordinates $(\rho,\phi)$, for which the metric takes the form (3.25). In these coordinates, the physical line element along a curve is $d\ell=e^{\varphi}\sqrt{d\rho^2+\rho^2 d\phi^2}$, and the relation \eqref{4.22} remains valid without modification. In particular, for a circular arc $\rho=\rho_\Gamma$ (so $d\rho=0$) with outward normal proportional to $\partial_\rho$, Eq. \eqref{4.22} reduces to an integral involving $\partial_\rho\varphi$ along the arc; this is precisely the mechanism by which the curvature area contribution is replaced by boundary data once the closing curve is chosen in the lensing domain.

\subsection{Boundary only expression for the deflection angle in isothermal coordinates} \label{sec4.3}
We now combine the finite distance Gauss-Bonnet formulation of Section \ref{sec2} with the isothermal identities of Section \ref{sec4.2} to obtain an expression for the deflection angle in which the curvature area term is replaced entirely by boundary data. In the static case, this viewpoint is closely aligned with the finite distance extension of the Gibbons-Werner method developed by Ishihara \textit{et al.} and related integral formulations.

Let $D\subset\mathcal{M}_{\rm opt}$ be a simply connected lensing domain with piecewise smooth boundary $\partial D$, chosen as in Section \ref{sec2.4} so that $\partial D$ contains the light ray segment $\gamma$ from the source $S$ to the receiver $R$ together with auxiliary boundary segments that close the contour. For a static optical geometry, the projected light ray is a geodesic of $(\mathcal{M}_{\rm opt},g_{ab})$, and hence its geodesic curvature vanishes as in Eq. \eqref{2.25}. If, in addition, the auxiliary radial segments are taken to be geodesics of the optical manifold (as in the standard construction in the finite distance literature), then the only nontrivial geodesic-curvature contribution comes from the closing segment $C_\Gamma\subset\partial D$. In that case, Gauss-Bonnet yields an integral representation of the form \cite{Allendoerfer_1943,Klingenberg2013,Carmo2016}
\begin{equation}
\alpha
= -\iint_D K\,dS \;-\; \int_{C_\Gamma} k_g\,d\ell \;+\; \Phi_{RS},
\label{4.23}
\end{equation}
where $\Phi_{RS}\equiv \phi_R-\phi_S$ is the coordinate separation angle introduced in Eq. \eqref{2.21}, and the Gauss-Bonnet corner-angle bookkeeping is chosen so that Eq. \eqref{4.23} reproduces the finite-distance deflection definition Eq. \eqref{2.22}.

We now rewrite the curvature area term in Eq. \eqref{4.23} using the isothermal reduction derived in Section \ref{sec4.2}. In an isothermal chart, Eq. \eqref{4.22} gives
\begin{equation}
\iint_D K\,dS
= -\oint_{\partial D} N^a\nabla_a\varphi\,d\ell,
\label{4.24}
\end{equation}
where $N^a$ is the outward unit normal to $\partial D$ with respect to the optical metric. Substituting Eq. \eqref{4.24} into Eq. \eqref{4.23}, we obtain the boundary-only representation
\begin{equation}
\alpha
=\oint_{\partial D} N^a\nabla_a\varphi\,d\ell \;-\; \int_{C_\Gamma} k_g\,d\ell \;+\; \Phi_{RS}.
\label{4.25}
\end{equation}
Since $\partial D$ is a union of smooth segments, Eq. \eqref{4.25} can be decomposed by writing $\partial D=\gamma\cup C_S\cup C_R\cup C_\Gamma$ with the induced orientation, giving
\begin{equation}
\alpha
=\int_{\gamma} N^a\nabla_a\varphi\,d\ell
+\int_{C_S} N^a\nabla_a\varphi\,d\ell
+\int_{C_R} N^a\nabla_a\varphi\,d\ell
+\int_{C_\Gamma}\left(N^a\nabla_a\varphi-k_g\right)d\ell
+\Phi_{RS}.
\label{4.26}
\end{equation}
Equation \eqref{4.26} is the desired structural statement: all dependence on the intrinsic curvature of the optical manifold has been traded for boundary evaluations of the conformal factor (through its normal derivative) and, possibly, the geodesic curvature of the chosen closing curve.

The usefulness of Eq. \eqref{4.26} becomes especially transparent in the axisymmetric isothermal polar coordinates $(\rho,\phi)$ introduced in Section \ref{sec3}, where \cite{Chern_1955,Gibbons:2008rj}
\begin{equation}
d\ell^2=e^{2\varphi(\rho)}\left(d\rho^2+\rho^2 d\phi^2\right).
\label{4.27}
\end{equation}
For a circular arc $C_\Gamma$ given by $\rho=\rho_\Gamma$ with $\phi$ as parameter, the outward unit normal is $N^a=e^{-\varphi},(\partial_\rho)^a$ and the line element is $d\ell=e^{\varphi}\rho_\Gamma\,d\phi$. Therefore,
\begin{equation}
N^a\nabla_a\varphi\,d\ell
=\left(e^{-\varphi}\partial_\rho\varphi\right)\left(e^{\varphi}\rho_\Gamma\,d\phi\right)
=\rho_\Gamma\,\partial_\rho\varphi(\rho_\Gamma)\,d\phi.
\label{4.28}
\end{equation}
If, moreover, $\varphi=\varphi(\rho)$ is axisymmetric, then the analogous contribution from a radial segment $\phi=\mathrm{const}$ vanishes because the normal derivative is proportional to $\partial_\phi\varphi=0$ along that segment. Under these common symmetry assumptions, Eq. \eqref{4.26} reduces to a sum of a light-ray boundary term and a closing-curve term, with the latter depending only on $\partial_\rho\varphi$ evaluated on $C_\Gamma$ and the geodesic curvature of $C_\Gamma$.

\section{Domain construction consistent with finite distance lensing} \label{sec5}
The boundary only representation derived in Section \ref{sec4} is only as useful as the choice of integration domain $D$: at finite source and receiver distances, one must construct a closed, oriented contour that (i) implements the finite distance definition of the deflection angle and (ii) yields boundary terms that are either computable directly or controllable in a weak-field expansion. We now specify a domain construction that is compatible with the definition \eqref{2.22} and the Gauss-Bonnet identity \eqref{2.23}, and that interfaces cleanly with the isothermal-coordinate framework of Section \ref{sec3} and Section \ref{sec4}.

\subsection{Finite distance lensing domain and boundary components} \label{sec5.1}
Let $\gamma\subset\mathcal{M}_{\rm opt}$ denote the spatial projection of the light ray on the equatorial optical manifold, connecting a source point $S$ to a receiver point $R$. We assume $S$ and $R$ lie in the region where the optical metric is Riemannian and where the isothermal construction is valid. The finite distance deflection angle is defined by the endpoint angles $\Psi_S$ and $\Psi_R$ through Eq. \eqref{2.22}, and Gauss-Bonnet will be applied to a simply connected region $D$ whose boundary contains $\gamma$ as one of its segments.

To close the contour at finite distance, we introduce two auxiliary curves $C_S$ and $C_R$ that emanate from the endpoints $S$ and $R$ and meet a third auxiliary curve $C_\Gamma$ that connects their other endpoints. Specifically, we define points $P_S$ and $P_R$ such that $P_S$ is connected to $S$ by $C_S$, $P_R$ is connected to $R$ by $C_R$, and $P_S$ is connected to $P_R$ by $C_\Gamma$. The boundary is then the positively oriented closed curve
\begin{equation}
\partial D=\gamma\cup C_R\cup C_\Gamma\cup C_S,
\label{5.1}
\end{equation}
where the segments are ordered so that $\partial D$ is traversed counterclockwise with respect to the chosen orientation on $\mathcal{M}_{\rm opt}$. The four corner points of $\partial D$ are $S$, $R$, $P_R$, and $P_S$, and their exterior angles enter Gauss-Bonnet through the corner sum in Eq. \eqref{2.23}.

The role of the curves $C_S$ and $C_R$ is to implement, geometrically, the local comparison between the light-ray direction and the radial direction that underlies the finite distance definition of deflection. In the polar-type chart $(r,\phi)$ used to define $\Phi_{RS}$ in Eq. \eqref{2.21}, we therefore choose $C_S$ and $C_R$ to be coordinate-radial curves of constant azimuth:
\begin{equation}
C_S:\ \phi=\phi_S,\qquad C_R:\ \phi=\phi_R,
\label{5.2}
\end{equation}
so that their tangents at $S$ and $R$ coincide with the outward radial unit direction used in Eq. \eqref{2.20}. This choice ensures that the endpoint angles entering Eq. \eqref{2.22} are precisely the corner-angle data associated with how $\gamma$ meets $C_S$ at $S$ and $C_R$ at $R$, once the orientation conventions are fixed. In the spherically symmetric static cases treated in the worked examples below, the equatorial optical metric takes the rotationally symmetric form
\begin{equation}
d\ell^2=A(r)\,dr^2+B(r)\,d\phi^2,
\label{5.3}
\end{equation}
so that the coordinate-radial curves $\phi=\mathrm{const}$ are optical geodesics: the $\phi$-geodesic equation implies $\frac{d}{d\lambda}\!\left(B(r)\,\dot\phi\right)=0$, and $\dot\phi\equiv 0$ solves it identically. Hence $k_g(C_S)=k_g(C_R)=0$ exactly for the choices in Eq. \eqref{5.2}. In more general (non-rotationally symmetric) optical geometries, the curves $C_S$ and $C_R$ need not be optical geodesics; in that case one must either (i) choose $C_S$ and $C_R$ to be true optical geodesics determined by the geodesic equations, or (ii) retain their geodesic-curvature contributions in the Gauss-Bonnet boundary sum.

The remaining curve $C_\Gamma$ is a closing segment whose main purpose is to render $\partial D$ closed while keeping its contributions under analytic control. In the isothermal polar coordinates \((\rho,\phi)\) introduced in Section \ref{sec3}, a natural and technically convenient option is to take \(C_\Gamma\) to be the \emph{circular arc} of the isothermal circular arc \(\rho=\rho_\Gamma\) that connects \(P_S\) and \(P_R\),
\begin{equation}
C_\Gamma:\ \rho=\rho_\Gamma,\qquad \phi\in[\phi_S,\phi_R]
\quad \text{(with orientation chosen to close } \partial D\text{)}.
\label{5.4}
\end{equation}
Thus, while the set \(\rho=\rho_\Gamma\) is a full circle in the \((u,v)\) plane, the boundary component entering \(\partial D\) is only the corresponding arc segment between the two auxiliary radial curves. This choice parallels the standard large-radius circular-arc closure used in asymptotic treatments, but here \(\rho_\Gamma\) is kept finite and is treated as a parameter specifying where the domain is closed.

\begin{figure}
    \centering
    \includegraphics[width=0.6\textwidth]{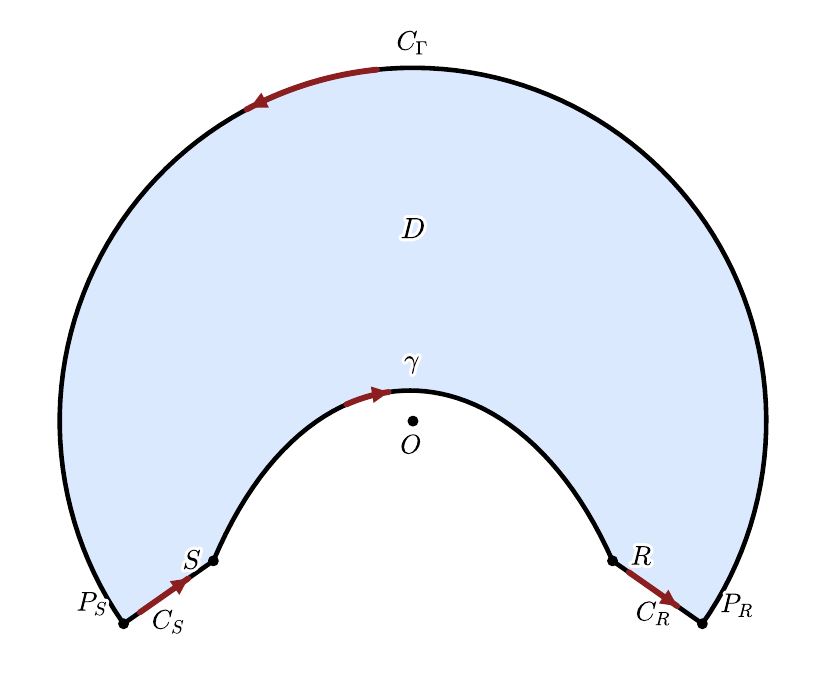}
    \caption{Schematic of the finite-distance Gauss--Bonnet integration domain on the equatorial optical manifold. The positively oriented boundary is \(\partial D=\gamma\cup C_R\cup C_\Gamma\cup C_S\).}
    \label{fig1}
\end{figure}

A schematic of the finite-distance Gauss--Bonnet domain and its positively oriented boundary is shown in Fig. \ref{fig1}.
With the boundary components fixed by Eqs. \eqref{5.1}-\eqref{5.4}, the deflection angle is represented by Gauss-Bonnet with explicit separation between (i) contributions tied to the physical light ray $\gamma$ and (ii) contributions tied to the auxiliary closure curves. The subsequent subsections are devoted to making this separation practically useful: we will specify closure choices for which the geodesic-curvature and normal-derivative terms simplify, and we will track the corner angles so that the resulting expression matches the finite distance definition \eqref{2.22} without ambiguity.

\subsection{Closing curve choices and eliminating/controlling geodesic curvature terms} \label{sec5.2}
The Gauss-Bonnet representation of the finite distance deflection angle depends on the geodesic curvature integrals along those boundary components of $\partial D$ that are not optical geodesics. In the boundary only formulation of Eq. \eqref{4.26}, this dependence is concentrated on the closing curve $C_\Gamma$ through the combination \(\left(N^a\nabla_a\varphi+k_g\right)d\ell\), together with any residual geodesic-curvature contributions on the remaining auxiliary segments if they are not chosen as geodesics. The practical objective in constructing $D$ is therefore to select boundary components whose geodesic curvature either vanishes identically or reduces to simple, controllable functions of the boundary coordinates.

We first note that the light-ray segment $\gamma$ is a geodesic of the optical metric in the static case, so \(k_g(\gamma)=0\) as already used in Eq. \eqref{2.25}. The same elimination principle may be applied to the radial segments \(C_S\) and \(C_R\): if these segments are chosen to be geodesics of \((\mathcal{M}_{\rm opt},g_{ab})\), then their geodesic curvature contributions vanish as well. While coordinate-radial curves \(\phi=\mathrm{const}\) in Eq. \eqref{5.2} need not be exact geodesics for a general axisymmetric metric, one can either (i) select geodesic segments that coincide with the required local radial direction at the endpoints, or (ii) retain their geodesic curvature contributions and treat them perturbatively in the weak-field regime. The first option is conceptually clean, while the second is often technically convenient when the coordinate choice is fixed by the underlying spacetime chart.

The closing curve \(C_\Gamma\) is the remaining essential ingredient. In isothermal coordinates, the optical metric is conformal to the flat metric, and this yields a simple relationship between the optical geodesic curvature \(k_g\) of a boundary curve and the corresponding Euclidean curvature. To make this explicit, we work in an isothermal chart \((u,v)\) in which
\begin{equation}
d\ell^2 = e^{2\varphi(u,v)}\left(du^2+dv^2\right).
\label{5.4b}
\end{equation}
Let \(C\subset\mathcal{M}_{\rm opt}\) be a smooth boundary segment, and let \(k_E\) denote its signed curvature with respect to the Euclidean metric \(du^2+dv^2\), with outward Euclidean unit normal \(\hat n^i\) and Euclidean arclength parameter \(s_E\). The corresponding optical geodesic curvature \(k_g\) and optical arclength \(d\ell\) satisfy the conformal transformation law
\begin{equation}
k_g = e^{-\varphi}\left(k_E+\hat n^i\partial_i\varphi\right),
\qquad
d\ell=e^{\varphi}\,ds_E.
\label{5.5}
\end{equation}
This identity isolates the nontrivial geometry into the normal derivative of \(\varphi\), while the Euclidean curvature \(k_E\) is determined purely by the shape of the closing curve in the \((u,v)\) plane.

Combining Eq. \eqref{5.5} with the normal-derivative term appearing in Eq. \eqref{4.26} yields a particularly convenient expression for the integrand along any smooth segment \(C\):
\begin{equation}
\left(N^a\nabla_a\varphi-k_g\right)d\ell
=\left(\hat n^i\partial_i\varphi\right)ds_E-\left(k_E+\hat n^i\partial_i\varphi\right)ds_E
=-k_E\,ds_E,
\label{5.6}
\end{equation}
where we used that, for a scalar, \(\nabla_a\varphi=\partial_a\varphi\) and that the optical normal derivative term reduces to \(\hat n^i\partial_i\varphi\,ds_E\) under the conformal rescalings.

Equation \eqref{5.6} shows that, in isothermal coordinates, the combination that actually enters the finite-distance Gauss-Bonnet deflection formula is independent of the conformal factor: it depends only on the Euclidean curvature $k_E$ of the chosen closing curve in the $(u,v)$ plane.

In the axisymmetric isothermal polar coordinates \((\rho,\phi)\) of Section \ref{sec3}, with
\begin{equation}
d\ell^2=e^{2\varphi(\rho)}\left(d\rho^2+\rho^2 d\phi^2\right),
\label{5.7}
\end{equation}
a natural option is the isothermal circular arc \(\rho=\rho_\Gamma\). For this curve, the Euclidean curvature is \(k_E=1/\rho_\Gamma\), the Euclidean arclength element is \(ds_E=\rho_\Gamma\,d\phi\), and the outward Euclidean unit normal is \(\hat n=\partial_\rho\). Substituting into Eq. \eqref{5.6} gives
\begin{equation}
\left(N^a\nabla_a\varphi-k_g\right)d\ell
=-k_E\,ds_E
=-\left(\frac{1}{\rho_\Gamma}\right)\left(\rho_\Gamma\,d\phi\right)
=-\,d\phi.
\label{5.8}
\end{equation}
In particular, for the isothermal circular arc closure $\rho=\rho_\Gamma$, Eq. \eqref{5.8} implies
\begin{equation}
\int_{C_\Gamma}\left(N^a\nabla_a\varphi-k_g\right)d\ell
=-\int_{\phi_S}^{\phi_R} d\phi
=-\Phi_{RS},
\end{equation}
so that the closure contribution cancels the explicit $\Phi_{RS}$ term in the finite-distance deflection angle.
\begin{figure}
    \centering
    \includegraphics[width=0.6\textwidth]{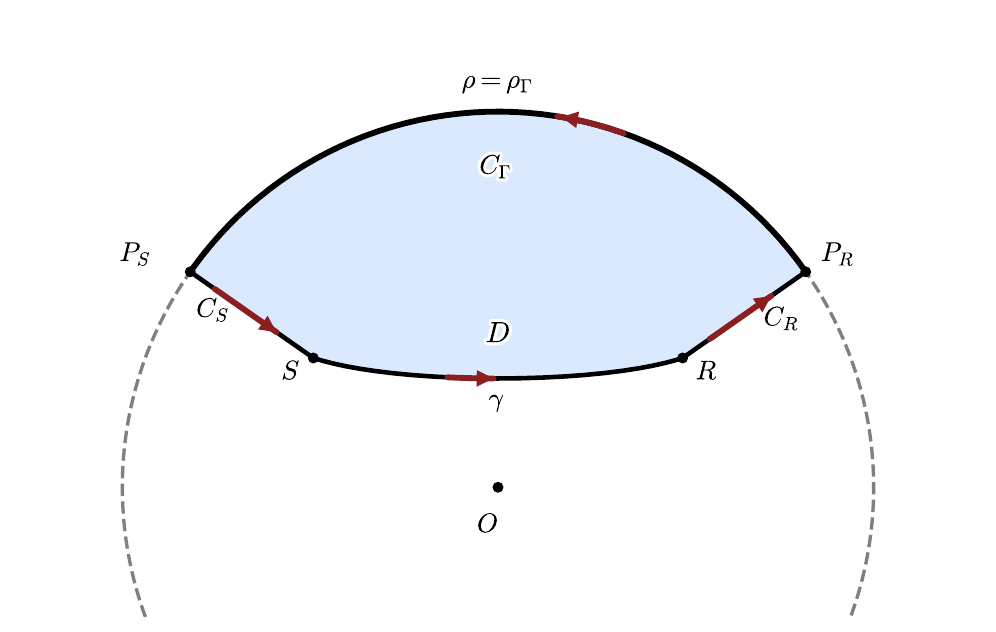}
    \caption{Closure at fixed isothermal radius. The dashed curve is the full constant-\(\rho\) circle \(\rho=\rho_\Gamma\), whereas the boundary component \(C_\Gamma\subset\partial D\) is only the solid circular arc between \(P_S\) and \(P_R\).}
    \label{fig2}
\end{figure}
Figure \ref{fig2} illustrates the special closure by a constant-isothermal-radius curve, \(\rho=\rho_\Gamma\). The dashed curve represents the full isothermal circle, while the actual boundary component entering \(\partial D\) is only the circular arc \(C_\Gamma\) joining \(P_S\) and \(P_R\). Together with the auxiliary radial segments \(C_S\) and \(C_R\) and the light-ray segment \(\gamma\), this arc encloses the domain \(D\). The figure makes clear that the Gauss--Bonnet construction depends only on this boundary arc, not on the entire circle.

Other closure choices are possible and can be advantageous depending on the finite distance configuration. For instance, if one chooses \(C_\Gamma\) so that its Euclidean curvature \(k_E\) is simple (piecewise constant, as for polygonal closures, or constant, as for circular arcs), then Eq. \eqref{5.6} shows that only the normal derivative of \(\varphi\) remains to be controlled. Conversely, if one can choose a closure curve whose Euclidean normal derivative of \(\varphi\) is small in the weak-field regime over the relevant range, then the leading contribution is governed by the purely Euclidean curvature term. In all cases, the guiding principle is that isothermal coordinates reduce the dependence on the bulk geometry of \(D\) to boundary evaluations of \(\varphi\) and its normal derivative, thereby making the finite distance closure analytically tractable.

\subsection{Corner angles, exterior-angle bookkeeping, and consistency with Ishihara-type definitions} \label{sec5.3}
To ensure that the Gauss-Bonnet construction reproduces the finite distance deflection angle defined in Eq. \eqref{2.22}, we must track carefully how the endpoint direction data \(\Psi_S\), \(\Psi_R\) and the coordinate separation \(\Phi_{RS}\) arise from the corner (exterior-angle) terms in Eq. \eqref{2.23}. This bookkeeping is the finite distance refinement of the original Gibbons-Werner construction and is standard in the Ishihara-type formulation.

We consider the quadrilateral-type boundary \(\partial D=\gamma\cup C_R\cup C_\Gamma\cup C_S\) introduced in Eq. \eqref{5.1}, oriented positively. The corner points are \(S\), \(R\), \(P_R\), and \(P_S\). Let \(\Theta_X\) denote the exterior (turning) angle at the corner point \(X\) as it appears in Gauss-Bonnet, i.e., the signed change in the boundary tangent direction when passing through \(X\) along \(\partial D\). With this convention, Eq. \eqref{2.23} reads
\begin{equation}
\iint_D K\,dS+\oint_{\partial D} k_g\,d\ell+\Theta_S+\Theta_R+\Theta_{P_R}+\Theta_{P_S}=2\pi.
\label{5.9}
\end{equation}

The endpoint angles \(\Psi_S\) and \(\Psi_R\) defined in Eq. \eqref{2.20} measure the optical angle between the light-ray tangent and the outward radial direction at \(S\) and \(R\). Because our boundary contains radial segments \(C_S\) and \(C_R\) aligned with the outward radial direction at the endpoints, these angles determine the turning at \(S\) and \(R\). With the orientation specified in Section \ref{sec5.1} (so that \(C_S\) is traversed inward toward \(S\) and \(C_R\) is traversed outward away from \(R\)), the exterior angles at the endpoints are
\begin{equation}
\Theta_S=\pi-\Psi_S,
\qquad
\Theta_R=\Psi_R.
\label{5.10}
\end{equation}
These relations are purely local and follow directly from comparing the oriented tangents of the adjacent boundary segments at each endpoint. They encode precisely the same direction comparison used to define \(\Psi_S\) and \(\Psi_R\), but now in the intrinsic optical geometry.

At the remaining corners \(P_S\) and \(P_R\), the radial segments meet the closing curve \(C_\Gamma\). A key simplification arises when \(C_\Gamma\) is chosen to be an isothermal circular arc \(\rho=\rho_\Gamma\) as in Eq. \eqref{5.3}. In the Euclidean reference metric \(d\rho^2+\rho^2 d\phi^2\), radial lines intersect circles orthogonally, and because the optical metric is conformal to this reference metric in isothermal coordinates, the right angles are preserved. Consequently, the interior angles at \(P_S\) and \(P_R\) are \(\pi/2\), and the corresponding exterior angles are also \(\pi/2\):
\begin{equation}
\Theta_{P_S}=\frac{\pi}{2},
\qquad
\Theta_{P_R}=\frac{\pi}{2}.
\label{5.11}
\end{equation}
Thus the two auxiliary corners contribute a universal amount \(\Theta_{P_S}+\Theta_{P_R}=\pi\), independent of the gravitational field, provided the closure is taken to be an isothermal circular arc and the radial segments are taken to be \(\phi=\) const.

Substituting Eqs. \eqref{5.10} and \eqref{5.11} into Eq. \eqref{5.9} yields
\begin{equation}
\iint_D K\,dS+\oint_{\partial D} k_g\,d\ell+\pi-\Psi_S+\Psi_R= \pi,
\label{5.12}
\end{equation}
or equivalently,
\begin{equation}
\Psi_R-\Psi_S = -\iint_D K\,dS-\oint_{\partial D} k_g\,d\ell.
\label{5.13}
\end{equation}

To connect Eq. \eqref{5.13} to the finite-distance deflection angle, we adopt the Ishihara-type definition
\begin{equation}
\alpha \equiv \Psi_R-\Psi_S+\Phi_{RS},
\label{5.14}
\end{equation}
where $\Psi_S$ and $\Psi_R$ are the endpoint angles defined in Eq. \eqref{2.21}, and $\Phi_{RS}\equiv \phi_R-\phi_S$ is the coordinate separation angle between the source and receiver.

Substituting Eq. \eqref{5.13} into Eq. \eqref{5.14} yields the explicit Gauss-Bonnet identity
\begin{equation}
\alpha
= -\iint_D K\,dS \;-\;\oint_{\partial D} k_g\,d\ell \;+\;\Phi_{RS}.
\label{5.15}
\end{equation}
Decomposing the boundary into the standard quadrilateral pieces,
\[
\partial D=\gamma\cup C_S\cup C_R\cup C_\Gamma,
\]
Eq. \eqref{5.15} becomes
\begin{equation}
\alpha
= -\iint_D K\,dS
-\int_{\gamma} k_g\,d\ell
-\int_{C_S} k_g\,d\ell
-\int_{C_R} k_g\,d\ell
-\int_{C_\Gamma} k_g\,d\ell
+\Phi_{RS}.
\label{5.16}
\end{equation}
For static spacetimes, the projected light ray is an optical geodesic, so $k_g(\gamma)=0$ (Eq. \eqref{2.25}). Moreover, one may choose the auxiliary curves $C_S$ and $C_R$ to be optical geodesics, so that their geodesic curvature terms vanish. Under these standard choices, Eq. \eqref{5.16} reduces to
\begin{equation}
\alpha
= -\iint_D K\,dS \;-\;\int_{C_\Gamma} k_g\,d\ell \;+\;\Phi_{RS},
\label{5.17}
\end{equation}
which is the finite-distance Gauss-Bonnet expression used in the subsequent isothermal-coordinate reduction.

\section{Practical normalization choices and avoidance of circular-orbit fixing} \label{sec6}
The isothermal-coordinate construction in Section \ref{sec3} determines the new radial coordinate \(\rho(r)\) only up to an overall multiplicative constant. This freedom is often fixed implicitly in the literature by appealing to a special geometric feature of the optical manifold (e.g., a circular photon orbit), but such a choice is neither necessary nor always convenient for finite distance lensing. We therefore isolate the normalization freedom at the level of the conformal factor and clarify which quantities are invariant under it.

\subsection{Additive freedom in the conformal factor and physical observables} \label{sec6.1}
In axisymmetric isothermal polar coordinates, the optical metric takes the form
\begin{equation}
d\ell^2=e^{2\varphi(\rho)}\left(d\rho^2+\rho^2 d\phi^2\right).
\label{6.1}
\end{equation}
As emphasized in Section \ref{sec3}, the radial map \(\rho(r)\) is determined by a first-order equation and hence contains one integration constant. Equivalently, if \(\rho(r)\) is one admissible isothermal radius, then so is the rescaled coordinate
\begin{equation}
\tilde\rho = c\,\rho,
\qquad c>0,
\label{6.2}
\end{equation}
with \(\phi\) unchanged. The same physical metric \(d\ell^2\) can be written in the same isothermal form in the \((\tilde\rho,\phi)\) chart, but with a shifted conformal factor. Indeed, substituting \(\rho=\tilde\rho/c\) and \(d\rho=d\tilde\rho/c\) into Eq. \eqref{6.1} gives
\begin{equation}
d\ell^2=e^{2\varphi(\tilde\rho/c)}\left(\frac{d\tilde\rho^2}{c^2}+\frac{\tilde\rho^2}{c^2}d\phi^2\right)
=e^{2\tilde\varphi(\tilde\rho)}\left(d\tilde\rho^2+\tilde\rho^2 d\phi^2\right),
\label{6.3}
\end{equation}
where the redefined conformal factor is
\begin{equation}
\tilde\varphi(\tilde\rho)=\varphi(\tilde\rho/c)-\ln c.
\label{6.4}
\end{equation}
Equations \eqref{6.2}-\eqref{6.4} show that the additive constant shift \(\varphi\mapsto\varphi-\ln c\) is not a physical rescaling of the optical metric; it is the compensating change induced by an allowed coordinate rescaling of the isothermal radius. The intrinsic geometry of \((\mathcal{M}_{\rm opt},g_{ab})\) is unchanged.

As a consequence, any physical observable constructed from the optical metric, such as endpoint angles defined by metric inner products, geodesic curvature computed with the Levi-Civita connection of \(g_{ab}\), or the Gaussian curvature \(K\), is independent of the choice of \(c\). In practical computations this invariance is most usefully tracked at the level of the boundary integrands that appear in the boundary only reduction.

For example, the curvature-to-boundary conversion in isothermal coordinates involves the normal derivative of \(\varphi\) (Section \ref{sec4.2}). Under the rescaling \eqref{6.2}-\eqref{6.4}, one finds
\begin{equation}
\partial_{\tilde\rho}\tilde\varphi(\tilde\rho)
=\frac{1}{c}\,\partial_{\rho}\varphi(\rho)\bigg|_{\rho=\tilde\rho/c},
\qquad
\tilde\rho\,\partial_{\tilde\rho}\tilde\varphi(\tilde\rho)
=\rho\,\partial_{\rho}\varphi(\rho).
\label{6.5}
\end{equation}
Thus, combinations that naturally occur on circular closures in isothermal polar form, specifically \(\rho\,\partial_\rho\varphi\), which enters the boundary expressions derived earlier, are invariant under the normalization choice. This is the operational statement that the integration constant in \(\rho(r)\) (and hence the additive constant in \(\varphi)\) is a coordinate gauge that does not affect the final deflection angle, provided all boundary terms are treated consistently in a single normalization.

\subsection{Asymptotic normalization in asymptotically flat spacetimes} \label{sec6.2}
When the underlying spacetime is asymptotically flat and static, it is natural to fix the residual scaling freedom in the isothermal radius by requiring that the optical metric approach the Euclidean metric at large distance from the lens. This choice ensures that the conformal factor \(\varphi\) measures only the gravitational deviation from flat propagation, rather than absorbing an arbitrary constant offset, and it aligns the finite distance construction smoothly with the usual asymptotic-deflection limit. Such a normalization is a convenient gauge choice available only in asymptotically flat cases; it is not part of the general boundary-only formalism developed here.

In the axisymmetric chart \((r,\phi)\) on the equatorial optical manifold, asymptotic flatness means that, as \(r\to\infty\), the optical metric coefficients satisfy
\begin{equation}
g_{rr}(r)=1+o(1),
\qquad
g_{\phi\phi}(r)=r^2\bigl(1+o(1)\bigr).
\label{6.6}
\end{equation}
Substituting Eq. \eqref{6.6} into the defining ODE for the isothermal radial map, Eq. \eqref{3.16}, yields the asymptotic behavior
\begin{equation}
\frac{d\rho}{dr}=\rho\,\sqrt{\frac{g_{rr}(r)}{g_{\phi\phi}(r)}}
=\frac{\rho}{r}\bigl(1+o(1)\bigr),
\qquad r\to\infty,
\label{6.7}
\end{equation}
so that \(\rho(r)\) grows linearly with \(r\) at large radius. The residual freedom \(\rho\mapsto c\rho\) is then fixed by imposing the asymptotic matching condition
\begin{equation}
\lim_{r\to\infty}\frac{\rho(r)}{r}=1.
\label{6.8}
\end{equation}
This condition is equivalent to choosing the multiplicative integration constant in Eq. \eqref{3.17} so that the isothermal radius coincides with the standard radial coordinate in the asymptotically flat region, up to subleading corrections determined by the gravitational field.

With Eq. \eqref{6.8} in place, the conformal factor in the isothermal polar form is asymptotically trivial. Using Eq. \eqref{3.22} together with Eq. \eqref{6.6} and Eq. \eqref{6.8}, we obtain
\begin{equation}
e^{2\varphi(\rho)}=\frac{g_{\phi\phi}(r(\rho))}{\rho^2}=1+o(1),
\qquad
\varphi(\rho)=o(1),
\qquad
\rho\to\infty.
\label{6.9}
\end{equation}
Thus the isothermal representation becomes canonically normalized: the optical metric approaches \(d\rho^2+\rho^2 d\phi^2\) in the asymptotic region, and the boundary integrals in the Gauss-Bonnet construction admit a clean separation into a flat reference contribution plus gravitational corrections controlled by derivatives of \(\varphi\). In particular, the boundary only curvature replacement depends on \(\nabla\varphi\) rather than on \(\varphi\) itself, so the asymptotic normalization eliminates an otherwise arbitrary constant while leaving all intrinsic observables unchanged as discussed in Section \ref{sec6.1}.

\subsection{Logarithmic conformal derivative and cancellation of flat-space contributions} \label{sec6.3}
In the boundary only representation of the deflection angle, the closure contribution along an isothermal circular arc naturally separates into a universal flat-space term and a curvature-induced correction. It is therefore useful to package the physically relevant part of the conformal factor into a normalization-independent quantity that vanishes in the flat limit and enters the boundary integrals without ambiguity.

We define the \emph{renormalized conformal factor} as the logarithmic radial derivative of \(\varphi\),
\begin{equation}
\psi(\rho)\equiv \frac{d\varphi}{d\ln\rho}=\rho\,\frac{d\varphi}{d\rho}.
\label{6.10}
\end{equation}
By construction, \(\psi\) is invariant under the residual rescaling \(\rho\mapsto c\rho\): under Eq. \eqref{6.2}-\eqref{6.4} the function \(\varphi\) shifts by a constant, but \(\rho\,\partial_\rho\varphi\) is unchanged, as already noted in Eq. \eqref{6.5}. Thus \(\psi\) captures precisely the coordinate-gauge-invariant content of the conformal factor relevant to lensing.

The connection to flat-space cancellation is immediate. For the isothermal circular closure \(\rho=\rho_\Gamma\), Eq. \eqref{5.8} gives
\begin{equation}
\left(N^a\nabla_a\varphi+k_g\right)d\ell
=\left(1+2\rho_\Gamma\,\partial_\rho\varphi(\rho_\Gamma)\right)d\phi
=\left(1+2\psi(\rho_\Gamma)\right)d\phi.
\label{6.11}
\end{equation}
The term \(1\,d\phi\) is the purely Euclidean turning contribution associated with the circular arc and is present even in flat space; the gravitational contribution is entirely contained in \(2\psi(\rho_\Gamma)\,d\phi\). Consequently, whenever the closure arc contributes through an integral over \(\phi\), one can remove the flat reference piece by subtracting the corresponding Euclidean arc contribution, leaving only the \(\psi\)-dependent correction. This implements a clean, scheme-independent cancellation of flat-space terms at finite distance.

It is also useful to express \(\psi\) directly in terms of the original axisymmetric optical-metric coefficients \(g_{rr}(r)\) and \(g_{\phi\phi}(r)\), without explicit reference to \(\rho(r)\). Using Eq. \eqref{3.24} and the identity
\begin{equation}
\frac{d\ln\rho}{dr}=\sqrt{\frac{g_{rr}(r)}{g_{\phi\phi}(r)}}
\label{6.12}
\end{equation}
from Eq. \eqref{3.16}, we obtain
\begin{equation}
\psi
=\frac{d\varphi/dr}{d\ln\rho/dr}
=\frac{\dfrac{1}{2}\dfrac{d}{dr}\ln g_{\phi\phi}(r)-\sqrt{\dfrac{g_{rr}(r)}{g_{\phi\phi}(r)}}}{\sqrt{\dfrac{g_{rr}(r)}{g_{\phi\phi}(r)}}}
=\frac{1}{2}\sqrt{\frac{g_{\phi\phi}(r)}{g_{rr}(r)}}\,\frac{d}{dr}\ln g_{\phi\phi}(r)-1,
\label{6.13}
\end{equation}
with the understanding that \(r=r(\rho)\) when \(\psi\) is regarded as a function of \(\rho\). In the Euclidean limit \(g_{rr}=1\) and \(g_{\phi\phi}=r^2\), the right-hand side of Eq. \eqref{6.13} vanishes identically, so \(\psi=0\) and the closure integrand in Eq. \eqref{6.11} reduces to \(d\phi\), as required.

We can see that \(\psi(\rho)\) is a convenient renormalized representative of the conformal factor: it is invariant under the additive freedom in \(\varphi\), it vanishes in flat space, and it appears linearly in the boundary only closure contribution, thereby isolating the genuinely gravitational part of the finite distance deflection angle.

\subsection{Weak-field expansion strategy and perturbative evaluation of the boundary integral} \label{sec6.4}
To implement the boundary only formalism in concrete lensing problems, we adopt a weak-field expansion in which the optical metric is written as a small deformation of the Euclidean metric in the asymptotic region. The practical point is that all contributions to the deflection angle can then be evaluated perturbatively on a reference (flat-space) domain, with the gravitational corrections entering only through explicit boundary data such as \(\psi=\rho\,\partial_\rho\varphi\) defined in Eq. \eqref{6.10}.

We assume an asymptotically flat, static configuration and introduce a formal bookkeeping parameter \(\varepsilon\) controlling the strength of the gravitational field in the domain relevant to the light ray. In the polar-type chart \((r,\phi)\) on \(\mathcal{M}_{\rm opt}\), we write the optical metric coefficients as
\begin{equation}
g_{rr}(r)=1+\varepsilon\,a_1(r)+\varepsilon^2 a_2(r)+O(\varepsilon^3),
\qquad
g_{\phi\phi}(r)=r^2\Bigl(1+\varepsilon\\,b_1(r)+\varepsilon^2 b_2(r)+O(\varepsilon^3)\Bigr),
\label{6.14}
\end{equation}
where \(a_n(r)\) and \(b_n(r)\) are dimensionless functions that decay at large \(r\) in a manner consistent with asymptotic flatness. Dimensional consistency is manifest because \(g_{rr}\) is dimensionless and \(g_{\phi\phi}\) carries the dimension of length squared.

The key boundary datum for the isothermal-circle closure is the logarithmic conformal derivative \(\psi\) from Eq. \eqref{6.10}.. Using the representation in terms of the original metric coefficients, Eq. \eqref{6.13}, and expanding Eq. \eqref{6.14} to first order, we obtain
\begin{equation}
\psi(r)=\varepsilon\,\psi_1(r)+O(\varepsilon^2),
\qquad
\psi_1(r)=\frac{1}{2}\Bigl(r\,b_1'(r)+b_1(r)-a_1(r)\Bigr),
\label{6.15}
\end{equation}
where a prime denotes \(d/dr\). Equation \eqref{6.15} shows explicitly that \(\psi\) is sourced by the mismatch between the angular and radial deformations of the optical metric, together with the radial variation of the angular deformation. In particular, \(\psi\to 0\) as \(r\to\infty\) whenever \(a_1\,b_1\to 0\) sufficiently fast.

To evaluate the deflection angle perturbatively, we proceed by expanding both the integrands and the integration curves around their flat-space counterparts. For a fixed impact parameter \(b\equiv L/E\) (with \(E\) and \(L\) defined in Eq. \eqref{2.8}), the flat reference trajectory in the Euclidean isothermal plane may be taken as a straight line at distance \(b\) from the origin. In Euclidean polar coordinates \((\rho,\phi)\) centered at the lens, this straight line can be written as
\begin{equation}
\rho_0(\phi)=\frac{b}{\sin\phi},
\label{6.16}
\end{equation}
where \(\phi=\pi/2\) corresponds to the point of closest approach. Finite source and receiver distances are implemented by endpoint angles \(\phi_S\) and \(\phi_R\) determined by the Euclidean relations \(\rho_S=b/\sin\phi_S\) and \(\rho_R=b/\sin\phi_R\), with \(\rho_S\) and \(\rho_R\) the isothermal radii of \(S\) and \(R\), respectively.

The boundary only representation derived earlier expresses \(\alpha\) as a sum of boundary integrals over segments of \(\partial D\). In the present setting, the dominant, explicitly controllable contribution arises from the isothermal-circle closure \(C_\Gamma:\rho=\rho_\Gamma\), for which Eq. \eqref{6.11} gives
\begin{equation}
\int_{C_\Gamma}\left(N^a\nabla_a\varphi+k_g\right)d\ell
=\int_{\phi_R}^{\phi_S}\left(1+2\psi(\rho_\Gamma)\right)d\phi,
\label{6.17}
\end{equation}
with the understanding that the limits are chosen consistently with the positive orientation of \(\partial D\). The flat-space term \(\int d\phi\) is universal and cancels against the corresponding purely Euclidean turning contribution implicit in the finite distance angle bookkeeping; the gravitational part of the closure is therefore isolated as
\begin{equation}
\delta_\Gamma \equiv 2\int_{\phi_R}^{\phi_S}\psi(\rho_\Gamma)\,d\phi.
\label{6.18}
\end{equation}
Using Eq. \eqref{6.15} and identifying \(\rho_\Gamma=r_\Gamma+O(\varepsilon)\) under the asymptotic normalization of Subsection \ref{6.2}, one obtains \(\delta_\Gamma\) to first order by evaluating \(\psi_1\) at the chosen closure radius and performing the elementary \(\phi\)-integration.

The remaining boundary contributions are treated by the same perturbative logic: one evaluates each segment integral on the zeroth-order domain and then includes corrections from (i) the first-order change in the integrand induced by \(a_1\,b_1\) and (ii) the first-order displacement of the curve relative to its flat reference shape. In practice, the calculation simplifies substantially under the axisymmetry assumption \(\varphi=\varphi(\rho)\). Then, along radial segments \(\phi=\mathrm{const}\), the normal derivative appearing in the curvature-replacement term depends only on \(\partial_\phi\varphi\), which vanishes identically, so these segments do not contribute to the \(N^a\nabla_a\varphi\) piece at any order in \(\varepsilon\). The light-ray segment \(\gamma\) is handled by evaluating all boundary data along the reference straight line \(\rho_0(\phi)\) in Eq. \eqref{6.16} and then adding the first-order correction from the gravitational bending of the trajectory; the latter can be organized consistently by writing \(\rho(\phi)=\rho_0(\phi)+\varepsilon\,\rho_1(\phi)+O(\varepsilon^2)\) and solving for \(\rho_1\) from the optical geodesic equation, or, equivalently, from the first integral associated with the conserved angular momentum in the optical metric.

Such a strategy yields a systematic expansion for \(\alpha\) in powers of \(\varepsilon\) in which every term is obtained from boundary evaluations of \(a_n(r)\), \(b_n(r)\), and their derivatives, together with elementary integrals over \(\phi\) on the reference domain. The essential technical advantage of the present formalism is that no bulk curvature integral is ever computed directly: the weak-field calculation is reduced to boundary data through \(\psi\) and the simple geometry of the chosen closure.

\section{Worked examples reproducing Li-type results} \label{sec7}
We now illustrate how the boundary only isothermal-coordinate framework developed in the preceding sections reproduces standard finite distance bending angles in canonical spacetimes. Our first example is the Schwarzschild geometry, where both the isothermal reduction and the weak-field finite distance deflection admit closed expressions that can be compared directly with the Gauss-Bonnet-based finite distance definitions in the literature. For each example below we explicitly provide (or summarize) the following ingredients so that the boundary-only evaluation can be reproduced without additional assumptions:
\begin{enumerate}
\item the spacetime lapse function (or metric functions) defining $g_{tt}$ and the induced equatorial optical metric $g_{ab}^{\rm opt}$;
\item the isothermal radial map $\rho(r)$ determined by Eq. \eqref{3.16} (including the choice of integration constants and the asymptotic/normalization convention);
\item the conformal factor $\varphi(\rho)$ entering $d\ell^2=e^{2\varphi}(d\rho^2+\rho^2d\phi^2)$;
\item the Gaussian curvature $K$ (or equivalently $\Delta\varphi$) and the resulting boundary representation of $\iint_D K\,dS$;
\item the closing curve choice $C_\Gamma$ and the explicit evaluation of its contribution (including the orientation and the cancellation with $\Phi_{RS}$ when applicable);
\item the weak-field/weak-deflection approximation used (e.g.\ straight-line zeroth-order ray) together with the order counting in $M/b$, $Q^2/b^2$, $\Lambda b^2$, etc.
\end{enumerate}

\subsection{Schwarzschild spacetime: isothermal reduction and finite distance deflection} \label{sec7.1}
We reconsider the Schwarzschild spacetime with mass parameter \(M>0\) in standard coordinates \((t,r,\theta,\phi)\) and signature \((-\,+,+,+)\). In units \(G=c=1\), the line element is
\begin{equation}
ds^2=-\left(1-\frac{2M}{r}\right)dt^2+\left(1-\frac{2M}{r}\right)^{-1}dr^2+r^2\left(d\theta^2+\sin^2\theta\,d\phi^2\right).
\label{7.1}
\end{equation}
Restricting to the equatorial plane \(\theta=\pi/2\), the two dimensional optical metric \(g_{ab}\) on \((r,\phi)\) is defined (static case) by dividing the spatial line element by \(-g_{tt}\). Writing \(f(r)\equiv 1-2M/r\), we obtain
\begin{equation}
d\ell^2=f(r)^{-2}\,dr^2+\frac{r^2}{f(r)}\,d\phi^2,
\qquad r>2M,
\label{7.2}
\end{equation}
which is Riemannian in the exterior region.

We now implement the isothermal-coordinate construction of Section \ref{sec3} and then compute the finite distance deflection angle using the boundary only Gauss-Bonnet reduction of Sections \ref{sec4}-\ref{sec5}.

We seek an isothermal polar radius \(\rho\) such that
\begin{equation}
d\ell^2=e^{2\varphi(\rho)}\left(d\rho^2+\rho^2\,d\phi^2\right),
\label{7.3}
\end{equation}
with \(\varphi\) depending only on \(\rho\) by spherical symmetry. Equating Eq. \eqref{7.2} with Eq. \eqref{7.3} and using the general ODE \(d\rho/\rho=\sqrt{g_{rr}/g_{\phi\phi}}\,dr\) yields
\begin{equation}
\frac{d\rho}{\rho}=\sqrt{\frac{f(r)^{-2}}{r^2/f(r)}}\,dr=\frac{dr}{r\sqrt{f(r)}}=\frac{dr}{r\sqrt{1-\frac{2M}{r}}}.
\label{7.4}
\end{equation}
Integrating Eq. \eqref{7.4} for \(r>2M\) and fixing the multiplicative constant by the asymptotic normalization \(\rho\sim r\) as \(r\to\infty\) gives the standard isotropic-radius relation
\begin{equation}
\rho(r)=\frac{1}{2}\left(r-M+\sqrt{r(r-2M)}\right),
\label{7.5}
\end{equation}
whose inverse is
\begin{equation}
r(\rho)=\rho\left(1+\frac{M}{2\rho}\right)^2.
\label{7.6}
\end{equation}
Substituting Eq. \eqref{7.6} into Eq. \eqref{7.2} yields the exact isothermal form \eqref{7.3} with
\begin{equation}
e^{2\varphi(\rho)}=\Omega(\rho)^2\equiv \frac{\left(1+\frac{M}{2\rho}\right)^6}{\left(1-\frac{M}{2\rho}\right)^2},
\qquad
\varphi(\rho)=3\ln\!\left(1+\frac{M}{2\rho}\right)-\ln\!\left(1-\frac{M}{2\rho}\right).
\label{7.7}
\end{equation}
This completes the isothermal reduction. We now compute the finite distance deflection angle \(\alpha\) in a way that explicitly exhibits the promised simplification: the bulk curvature integral is never evaluated. Instead, we reduce \(\alpha\) to a single boundary integral along the light ray in the \((u,v)\) plane.

In the finite distance Gauss-Bonnet construction of Ishihara \textit{et al.}, the bending angle is defined by endpoint direction comparisons and an azimuthal separation, \(\alpha=\Psi_R-\Psi_S+\Phi_{RS}\), and is related to Gauss-Bonnet by applying the theorem to a quadrilateral-like domain \(D\) bounded by the light ray, two radial lines, and a closing curve. We take precisely such a domain on the optical manifold written in isothermal polar coordinates \((\rho,\phi)\), and we choose the closing curve to be an isothermal circular arc \(\rho=\rho_\Gamma\). This choice ensures orthogonality between the circle and the radial lines in the conformal (isothermal) geometry, so that the auxiliary corner exterior angles are \(\pi/2\) each, as used in Section \ref{sec5}.

For this domain, Gauss-Bonnet gives
\begin{equation}
\iint_D K\,dS+\int_{\partial D}k_g\,d\ell+\sum_{i}\Theta_i=2\pi,
\label{7.8}
\end{equation}
where \(K\) is the Gaussian curvature of the optical metric, \(k_g\) is the geodesic curvature of the boundary segments, and \(\Theta_i\) are the exterior angles at the corners. In the static Schwarzschild case, the light ray segment \(\gamma\) is an optical geodesic, so \(k_g(\gamma)=0\). Taking the radial segments to be optical geodesics (or, equivalently for the present weak-field calculation, retaining only their vanishing leading contribution under axisymmetry) leaves only the closing circular arc \(C_\Gamma\) as a possible source of nonzero \(k_g\). The corner-angle bookkeeping for this domain yields the identity
\begin{equation}
\alpha = -\iint_D K\,dS-\int_{C_\Gamma} k_g\,d\ell+\Phi_{RS},
\label{7.9}
\end{equation}
with \(\Phi_{RS}\equiv \phi_R-\phi_S\). The crucial point of Eq. \eqref{7.9} is the explicit subtraction of the closure turning contribution: in flat space \((M=0)\), one has \(K=0\) and \(k_g\,d\ell=d\phi\) on a Euclidean circle, so Eq. \eqref{7.9} gives \(\alpha=0\) identically.

We now eliminate the curvature area term using the isothermal identity \(K=-\Delta\varphi\) derived in Section \ref{sec4}. In isothermal coordinates, the curvature area integral reduces to a boundary term,
\begin{equation}
\iint_D K\,dS=-\oint_{\partial D} N^a\nabla_a\varphi\,d\ell,
\label{7.10}
\end{equation}
where \(N^a\) is the outward unit normal to \(\partial D\) in the optical metric. Substituting Eq. \eqref{7.10} into Eq. \eqref{7.9} gives
\begin{equation}
\alpha=\oint_{\partial D} N^a\nabla_a\varphi\,d\ell-\int_{C_\Gamma} k_g\,d\ell+\Phi_{RS}.
\label{7.11}
\end{equation}
We now evaluate the \(C_\Gamma\) contribution explicitly and show that it cancels \(\Phi_{RS}\) identically, leaving a single integral along the light ray. On the isothermal circular arc \(C_\Gamma:\rho=\rho_\Gamma\), we have \(d\ell=e^{\varphi}\rho_\Gamma\,d\phi\) and \(N^a\nabla_a\varphi=\partial_\rho\varphi/e^{\varphi}\). Hence
\begin{equation}
\left(N^a\nabla_a\varphi\right)d\ell
=\rho_\Gamma\,\partial_\rho\varphi(\rho_\Gamma)\,d\phi.
\label{7.12}
\end{equation}
The geodesic curvature of \(C_\Gamma\) in the optical metric can be obtained from the conformal transformation law \(k_g=e^{-\varphi}(k_E+\partial_n\varphi)\), where \(k_E=1/\rho_\Gamma\) is the Euclidean curvature of the circle and \(\partial_n\varphi=\partial_\rho\varphi\). This gives
\begin{equation}
k_g\,d\ell=(k_E+\partial_\rho\varphi)\,ds_E=\left(\frac{1}{\rho_\Gamma}+\partial_\rho\varphi(\rho_\Gamma)\right)\rho_\Gamma\,d\phi
=\left(1+\rho_\Gamma\,\partial_\rho\varphi(\rho_\Gamma)\right)d\phi.
\label{7.13}
\end{equation}
Subtracting Eq. \eqref{7.13} from Eq. \eqref{7.12} yields the identity
\begin{equation}
\left[\left(N^a\nabla_a\varphi\right)d\ell-k_g\,d\ell\right]_{C_\Gamma}=-\,d\phi.
\label{7.14}
\end{equation}
Therefore, upon integrating along \(C_\Gamma\) from \(\phi_S\) to \(\phi_R\),
\begin{equation}
\int_{C_\Gamma}\left[\left(N^a\nabla_a\varphi\right)d\ell-k_g\,d\ell\right]=-\int_{\phi_S}^{\phi_R}d\phi=-\Phi_{RS}.
\label{7.15}
\end{equation}
Substituting Eq. \eqref{7.15} into Eq. \eqref{7.11} shows that the entire closure contribution cancels the explicit \(\Phi_{RS}\) term:
\begin{equation}
\alpha=\int_{\gamma} N^a\nabla_a\varphi\,d\ell,
\label{7.16}
\end{equation}
where we have used that the axisymmetry \(\varphi=\varphi(\rho)\) implies the radial segments contribute nothing to \(\oint_{\partial D} N^a\nabla_a\varphi\,d\ell\) (their outward normal is proportional to \(\partial_\phi\), while \(\partial_\phi\varphi=0)\). Equation \eqref{7.16} is the promised simplification for Schwarzschild with an isothermal-circular closure: the deflection angle reduces to a single boundary integral along the light ray, with no bulk curvature integral and no orbit dependent normalization such as a circular photon orbit.

To evaluate Eq. \eqref{7.16} in the weak-field regime, we pass to isothermal Cartesian coordinates \((u,v)\) defined by
\begin{equation}
u=\rho\cos\phi,\qquad v=\rho\sin\phi,
\label{7.17}
\end{equation}
so that \(d\rho^2+\rho^2 d\phi^2=du^2+dv^2\). The optical metric is \(d\ell^2=e^{2\varphi(u,v)}(du^2+dv^2)\), and the key identity from Section \ref{sec4.2} implies that along any boundary segment,
\begin{equation}
N^a\nabla_a\varphi\,d\ell=\hat n^i\partial_i\varphi\,ds_E,
\label{7.18}
\end{equation}
where \(\hat n^i\) and \(ds_E\) are the Euclidean outward unit normal and Euclidean line element in the \((u,v)\) plane. Thus Eq. \eqref{7.16} becomes
\begin{equation}
\alpha=\int_{\gamma}\hat n^i\partial_i\varphi\,ds_E.
\label{7.19}
\end{equation}

In the weak-field regime, we evaluate Eq. \eqref{7.19} on the zeroth-order (flat) ray, which is a straight line at Euclidean impact parameter \(b\) from the origin. We choose coordinates so that this line is \(v=b\) with \(u\) increasing from the source to the receiver. The domain \(D\) is taken to lie on the side \(v\ge b\), so that the outward normal along \(\gamma\) points in the negative \(v\) direction, \(\hat n=-\partial_v\), and \(ds_E=du\). Since \(\rho=\sqrt{u^2+v^2}\), we have
\begin{equation}
\hat n^i\partial_i\varphi=-\partial_v\varphi(\rho)
=-\varphi'(\rho)\frac{v}{\rho}
=-\varphi'(\rho)\frac{b}{\rho}.
\label{7.20}
\end{equation}
From the exact expression \eqref{7.7}, the leading weak-field expansion at large \(\rho\) is
\begin{equation}
\varphi(\rho)=\frac{2M}{\rho}+O\!\left(\frac{M^2}{\rho^2}\right),
\qquad
\varphi'(\rho)=-\frac{2M}{\rho^2}+O\!\left(\frac{M^2}{\rho^3}\right).
\label{7.21}
\end{equation}
Substituting Eq. \eqref{7.21} into Eq. \eqref{7.20} and then into Eq. \eqref{7.19} yields, to first order in \(M\),
\begin{equation}
\alpha = \int_{u_S}^{u_R}\frac{2Mb}{\rho^3}\,du+O(M^2),
\qquad
\rho=\sqrt{u^2+b^2},
\label{7.22}
\end{equation}
where \(u_S<0<u_R\) correspond to the finite source and receiver locations on the line \(v=b\). The remaining integral is elementary:
\begin{equation}
\int \frac{du}{(u^2+b^2)^{3/2}}=\frac{u}{b^2\sqrt{u^2+b^2}},
\label{7.23}
\end{equation}
so Eq. \eqref{7.22} gives
\begin{equation}
\alpha
=\frac{2M}{b}\left[\frac{u}{\sqrt{u^2+b^2}}\right]_{u_S}^{u_R}+O(M^2).
\label{7.24}
\end{equation}
Define \(\rho_S\equiv\sqrt{u_S^2+b^2}\) and \(\rho_R\equiv\sqrt{u_R^2+b^2}\), the Euclidean (isothermal) radii of the source and receiver. Then \(u_R/\rho_R=+\sqrt{1-b^2/\rho_R^2}\) and \(u_S/\rho_S=-\sqrt{1-b^2/\rho_S^2}\), so Eq. \eqref{7.24} becomes
\begin{equation}
\alpha
=\frac{2M}{b}\left(\sqrt{1-\frac{b^2}{\rho_R^2}}+\sqrt{1-\frac{b^2}{\rho_S^2}}\right)+O(M^2).
\label{7.25}
\end{equation}
Finally, under the asymptotic normalization of Section \ref{sec6.2}, \(\rho=r+O(M)\) in the weak-field region, so to first order we may replace \(\rho_S\) and \(\rho_R\) by the usual areal radii \(r_S\) and \(r_R\). Writing \(u_S\equiv 1/r_S\) and \(u_R\equiv 1/r_R\), we obtain the standard finite distance weak-deflection result
\begin{equation}
\alpha
=\frac{2M}{b}\left(\sqrt{1-b^2u_R^2}+\sqrt{1-b^2u_S^2}\right)+O(M^2),
\label{7.26}
\end{equation}
which reduces to \(\alpha\to 4M/b\) in the limit \(r_S,r_R\to\infty\). This agrees with the finite distance Gauss-Bonnet definition and its Schwarzschild evaluation in the literature.

Equation \eqref{7.16} is the core showcase of the formalism: once the Schwarzschild optical metric is put in isothermal form, the finite distance deflection angle at leading order is obtained by a single normal-derivative integral along the (flat) reference ray, with the closure and azimuthal bookkeeping canceling identically at the integrand level via Eq. \eqref{7.14}.

\subsection{Reissner-Nordstr\"om spacetime: boundary only evaluation and comparison} \label{sec7.2}
We next consider the Reissner-Nordstr\"om spacetime with mass parameter \(M\) and electric charge parameter \(Q\) in units \(G=c=1\). In standard coordinates \((t,r,\theta,\phi)\), the line element is
\begin{equation}
ds^2=-f(r)\,dt^2+f(r)^{-1}\,dr^2+r^2\left(d\theta^2+\sin^2\theta\,d\phi^2\right),
\qquad
f(r)\equiv 1-\frac{2M}{r}+\frac{Q^2}{r^2}.
\label{7.27}
\end{equation}
Restricting to the equatorial plane \(\theta=\pi/2\) and using the static optical-metric construction, the two dimensional equatorial optical metric is
\begin{equation}
d\ell^2=f(r)^{-2}\,dr^2+\frac{r^2}{f(r)}\,d\phi^2,
\label{7.28}
\end{equation}
valid in the exterior region where the optical metric is Riemannian.

Our goal is to compute the finite distance weak deflection angle \(\alpha\) using the boundary only formula exhibited in Section \ref{sec7.1}. We therefore adopt the same Gauss-Bonnet domain as in Section \ref{sec5} (light ray \(\gamma\), two radial segments, and an isothermal circular closure \(C_\Gamma)\), and we use the same cancellation mechanism: for an isothermal circular arc closure, the closure contribution cancels \(\Phi_{RS}\) identically, leaving
\begin{equation}
\alpha=\int_{\gamma} N^a\nabla_a\varphi\,d\ell
=\int_{\gamma}\hat n^{,i}\,\partial_i\varphi\,ds_E,
\label{7.29}
\end{equation}
where \(\varphi\) is the conformal factor in isothermal coordinates, \(\hat n^{\,i}\) and \(ds_E\) are respectively the Euclidean unit normal and Euclidean arclength element in the isothermal \((u,v)\) plane, and we used the conformal identity of Eq. \eqref{7.18} established in Section \ref{7.1}. This reduction is exactly the point that we need to show: no bulk curvature integral is evaluated, and no orbit dependent fixing (e.g., via a photon sphere) is used. The finite distance definition itself is the Ishihara \emph{et al.} Gauss-Bonnet definition.

To evaluate Eq. \eqref{7.29} in the weak-field regime, we require the large-radius expansion of \(\varphi\). We proceed in two steps: (i) construct an isothermal radius \(\rho\) asymptotically, and (ii) reconstruct \(\varphi\) from the angular coefficient.

From Section \ref{sec3}, the isothermal polar ansatz \(d\ell^2=e^{2\varphi(\rho)}(d\rho^2+\rho^2 d\phi^2)\) implies the radial ODE
\begin{equation}
\frac{d\rho}{\rho}=\sqrt{\frac{g_{rr}(r)}{g_{\phi\phi}(r)}}\,dr
=\frac{dr}{r\sqrt{f(r)}}.
\label{7.30}
\end{equation}
In the weak-field region \(r\gg M\) with \(Q^2/r^2\ll 1\), we expand
\begin{equation}
\frac{1}{\sqrt{f(r)}} = 1+\frac{M}{r}+\frac{3M^2-Q^2}{2r^2}+O(r^{-3}),
\label{7.31}
\end{equation}
so Eq. \eqref{7.30} integrates to
\begin{equation}
\ln\rho=\ln r-\frac{M}{r}-\frac{3M^2-Q^2}{4r^2}+O(r^{-3}).
\label{7.32}
\end{equation}
Fixing the multiplicative integration constant by asymptotic normalization \(\rho/r\to 1\) as \(r\to\infty\), we obtain
\begin{equation}
\rho=r-M+\frac{Q^2-M^2}{4r}+O(r^{-2}),
\qquad
r=\rho+M+\frac{M^2-Q^2}{4\rho}+O(\rho^{-2}).
\label{7.33}
\end{equation}

Next we reconstruct the conformal factor from \(g_{\phi\phi}\) using Eq. \eqref{3.22},
\begin{equation}
e^{2\varphi(\rho)}=\frac{g_{\phi\phi}(r(\rho))}{\rho^2}
=\frac{r(\rho)^2}{f(r(\rho))\,\rho^2}.
\label{7.34}
\end{equation}
Substituting Eq. \eqref{7.33} and expanding for large \(\rho\) yields
\begin{equation}
e^{2\varphi(\rho)}=1+\frac{4M}{\rho}+\frac{15M^2-3Q^2}{2\rho^2}+O(\rho^{-3}).
\label{7.35}
\end{equation}
Taking the logarithm, we find
\begin{equation}
\varphi(\rho)=\frac{2M}{\rho}-\frac{M^2+3Q^2}{4\rho^2}+O(\rho^{-3}).
\label{7.36}
\end{equation}
For the weak-deflection calculation below we retain the leading mass term \(O(M/\rho)\) and the leading charge correction \(O(Q^2/\rho^2)\). We do not attempt to capture the full \(O(M^2)\) bending in the final \(\alpha\), because that requires consistently including the \(O(M)\) perturbation of the ray shape (and related second-order boundary corrections) beyond the straight-line evaluation used here; by contrast, the \(Q^2\) term we extract below already appears at order \(1/b^2\) with a universal coefficient in the asymptotic limit and is cleanly accessible within the present boundary evaluation. The asymptotic example for this coefficient is well known, including in Gauss-Bonnet computations for Reissner-Nordstr\"om. 

We now evaluate Eq. \eqref{7.29} in isothermal Cartesian coordinates \((u,v)\), related to \((\rho,\phi)\) by \(u=\rho\cos\phi\), \(v=\rho\sin\phi\). In the weak field we evaluate the boundary integral on the zeroth-order ray, taken to be the straight line
\begin{equation}
v=b,
\qquad
u\in[u_S,u_R],
\label{7.37}
\end{equation}
where \(b>0\) is the Euclidean impact parameter in the \((u,v)\) plane and \(u_S<0<u_R\) correspond to the finite source and receiver locations. Along this line, \(\rho=\sqrt{u^2+b^2}\), \(ds_E=du\), and (with the domain chosen on the side \(v\ge b\) as in Section \ref{sec7.1}) the outward Euclidean normal is \(\hat n=-\partial_v\). Therefore,
\begin{equation}
\hat n^{,i}\partial_i\varphi=-\partial_v\varphi(\rho)
=-\varphi'(\rho)\frac{\partial\rho}{\partial v}
=-\varphi'(\rho)\frac{b}{\rho},
\label{7.38}
\end{equation}
and Eq. \eqref{7.29} becomes
\begin{equation}
\alpha=\int_{u_S}^{u_R}\left[-\varphi'(\rho)\frac{b}{\rho}\right]du.
\label{7.39}
\end{equation}
Differentiating Eq. \eqref{7.36} gives
\begin{equation}
\varphi'(\rho)=-\frac{2M}{\rho^2}+\frac{M^2+3Q^2}{2\rho^3}+O(\rho^{-4}).
\label{7.40}
\end{equation}
Keeping the terms needed to reproduce the leading mass bending and the leading charge correction, Eq. \eqref{7.39} becomes
\begin{equation}
\alpha=\int_{u_S}^{u_R}\left(\frac{2Mb}{\rho^3}-\frac{3Q^2\,b}{2\rho^4}\right)du
+O\!\left(\frac{M^2}{b^2}\,\frac{MQ^2}{b^3}\,\frac{Q^4}{b^4}\right),
\qquad
\rho=\sqrt{u^2+b^2}.
\label{7.41}
\end{equation}

The first integral is identical to the Schwarzschild case:
\begin{equation}
\int \frac{du}{(u^2+b^2)^{3/2}}=\frac{u}{b^2\sqrt{u^2+b^2}},
\label{7.42}
\end{equation}
so the mass contribution is
\begin{equation}
\alpha_M=\frac{2M}{b}\left[\frac{u}{\sqrt{u^2+b^2}}\right]_{u_S}^{u_R}.
\label{7.43}
\end{equation}
Define the Euclidean (isothermal) endpoint radii \(\rho_R=\sqrt{u_R^2+b^2}\) and \(\rho_S=\sqrt{u_S^2+b^2}\). Then \(u_R/\rho_R=+\sqrt{1-b^2/\rho_R^2}\) and \(u_S/\rho_S=-\sqrt{1-b^2/\rho_S^2}\), giving
\begin{equation}
\alpha_M=\frac{2M}{b}\left(\sqrt{1-\frac{b^2}{\rho_R^2}}+\sqrt{1-\frac{b^2}{\rho_S^2}}\right).
\label{7.44}
\end{equation}

For the charge correction we use the elementary antiderivative
\begin{equation}
\int \frac{du}{(u^2+b^2)^2}=\frac{u}{2b^2(u^2+b^2)}+\frac{1}{2b^3}\arctan\!\left(\frac{u}{b}\right),
\label{7.45}
\end{equation}
so that
\begin{equation}
\alpha_Q
=-\frac{3Q^2,b}{2}\left[\frac{u}{2b^2(u^2+b^2)}+\frac{1}{2b^3}\arctan\!\left(\frac{u}{b}\right)\right]_{u_S}^{u_R}
=-\frac{3Q^2}{4}\left[\frac{u}{b(u^2+b^2)}+\frac{1}{b^2}\arctan\!\left(\frac{u}{b}\right)\right]_{u_S}^{u_R}.
\label{7.46}
\end{equation}
The first bracketed term satisfies \(u/(u^2+b^2)=u/\rho^2\), hence
\begin{equation}
\left[\frac{u}{u^2+b^2}\right]_{u_S}^{u_R}
=\frac{u_R}{\rho_R^2}-\frac{u_S}{\rho_S^2}
=\frac{1}{\rho_R}\sqrt{1-\frac{b^2}{\rho_R^2}}+\frac{1}{\rho_S}\sqrt{1-\frac{b^2}{\rho_S^2}}.
\label{7.47}
\end{equation}
For the arctangent term, introduce angles \(\theta_R,\theta_S\in(0,\pi/2)\) via \(\sin\theta_a=b/\rho_a\) at \(a\in{R,S}\). Then for \(u_R>0\) and \(u_S<0\),
\begin{equation}
\arctan\!\left(\frac{u_R}{b}\right)=\frac{\pi}{2}-\theta_R,
\qquad
\arctan\!\left(\frac{u_S}{b}\right)=-\frac{\pi}{2}+\theta_S,
\label{7.48}
\end{equation}
so
\begin{equation}
\left[\arctan\!\left(\frac{u}{b}\right)\right]_{u_S}^{u_R}
=\pi-(\theta_R+\theta_S)
=\pi-\left[\arcsin\!\left(\frac{b}{\rho_R}\right)+\arcsin\!\left(\frac{b}{\rho_S}\right)\right].
\label{7.49}
\end{equation}
Substituting Eqs. \eqref{7.47}-\eqref{7.49} into Eq. \eqref{7.46} yields the explicit finite distance charge correction
\begin{equation}
\alpha_Q
=-\frac{3Q^2}{4b^2}\left[
\pi-\arcsin\!\left(\frac{b}{\rho_R}\right)-\arcsin\!\left(\frac{b}{\rho_S}\right)
+\frac{b}{\rho_R}\sqrt{1-\frac{b^2}{\rho_R^2}}
+\frac{b}{\rho_S}\sqrt{1-\frac{b^2}{\rho_S^2}}
\right].
\label{7.50}
\end{equation}
Combining Eqs. \eqref{7.44} and \eqref{7.50}, we arrive at the weak-deflection finite distance result
\begin{eqnarray}
\alpha
=\frac{2M}{b}\left(\sqrt{1-\frac{b^2}{\rho_R^2}}+\sqrt{1-\frac{b^2}{\rho_S^2}}\right)
-\frac{3Q^2}{4b^2}\left[
\pi-\arcsin\!\left(\frac{b}{\rho_R}\right)-\arcsin\!\left(\frac{b}{\rho_S}\right)
+\frac{b}{\rho_R}\sqrt{1-\frac{b^2}{\rho_R^2}}
+\frac{b}{\rho_S}\sqrt{1-\frac{b^2}{\rho_S^2}}
\right] \notag
\\+O\!\left(\frac{M^2}{b^2}\,\frac{MQ^2}{b^3}\,\frac{Q^4}{b^4}\right).
\label{7.51}
\end{eqnarray}
Under the asymptotic normalization of Section \ref{sec6.2}, \(\rho=r+O(M,Q^2/r)\), so at this order we may replace \(\rho_S\,\rho_R\) by the areal radii \(r_S,r_R\).

 In the infinite distance limit \(r_S,r_R\to\infty\), Eq. \eqref{7.51} reduces to
\begin{equation}
\alpha \to \frac{4M}{b}-\frac{3\pi Q^2}{4b^2}+O\!\left(\frac{M^2}{b^2}\,\frac{MQ^2}{b^3}\,\frac{Q^4}{b^4}\right),
\label{7.52}
\end{equation}
in agreement with Gauss-Bonnet computations for Reissner-Nordstr\"om in the weak limit.  Moreover, at \(Q=0\), Eq. \eqref{7.51} reduces to the Schwarzschild finite distance expression already derived in Eq. \eqref{7.25}.

From the perspective of the present work, the operational simplification is visible in the structure of the calculation: once Eq. \eqref{7.29} is established, the entire weak-field bending computation reduces to (i) an asymptotic expansion of the conformal factor \(\varphi(\rho)\) and (ii) one-dimensional Euclidean integrals along the straight reference ray, with the finite distance dependence entering only through the finite endpoints \(\rho_S\,\rho_R\) and the elementary functions in Eq. \eqref{7.50}.

\subsection{Kottler (Schwarzschild-de Sitter) spacetime: total finite-distance deflection angle} \label{sec7.3}
In this subsection we re-derive the \emph{total} finite-distance deflection angle
\begin{equation}
\alpha \equiv \Psi_R-\Psi_S+\Phi_{RS},
\label{eq:alpha_def_Kottler}
\end{equation}
for the Kottler spacetime using the boundary-only isothermal formalism. The purpose is twofold:
(i) to demonstrate that the present method applies beyond asymptotically flat geometries, and
(ii) to provide a transparent comparison with the standard finite-distance result (in particular, the explicit $\mathcal{O}(\Lambda)$ and $\mathcal{O}(\Lambda M)$ terms).

The Kottler metric is given by
\begin{equation}
ds^2=-f(r)\,dt^2+\frac{dr^2}{f(r)}+r^2\left(d\theta^2+\sin^2\theta\,d\phi^2\right),
\qquad
f(r)=1-\frac{2M}{r}-\frac{\Lambda}{3}r^2.
\label{eq:Kottler_metric}
\end{equation}
Restricting to the equatorial plane $\theta=\pi/2$, the corresponding 2D optical metric is
\begin{equation}
d\ell^2 = \frac{dr^2}{f(r)^2}+\frac{r^2}{f(r)}\,d\phi^2.
\label{eq:Kottler_optical}
\end{equation}

Let $(u,v)$ be isothermal coordinates on the optical manifold, with polar form
\begin{equation}
u=\rho\cos\phi,\qquad v=\rho\sin\phi,\qquad ds_E^2\equiv du^2+dv^2=d\rho^2+\rho^2d\phi^2,
\label{eq:isothermal_plane}
\end{equation}
such that
\begin{equation}
d\ell^2 = e^{2\varphi(\rho)}\,ds_E^2.
\label{eq:isothermal_conformal}
\end{equation}
For the standard finite-distance quadrilateral domain $D$ (bounded by the photon curve $\gamma$, two auxiliary optical geodesics $C_S,C_R$, and an isothermal circular arc $C_\Gamma:\rho=\rho_\Gamma$), the Gauss-Bonnet formula in the static case reduces to the boundary-only expression
\begin{equation}
\alpha = \int_{\gamma} N^a\nabla_a\varphi\,d\ell,
\label{eq:alpha_boundary_only}
\end{equation}
where $N^a$ is the outward unit normal to $\gamma$ with respect to the optical metric. (The contributions from $C_\Gamma$ cancel the explicit $\Phi_{RS}$ term for the circular closure, while $k_g(\gamma)=0$ and $C_S,C_R$ may be chosen as optical geodesics.)

Using $d\ell=e^{\varphi}ds_E$ and $N^a\nabla_a\varphi\,d\ell=\hat n^{\,i}\partial_i\varphi\,ds_E$ in the isothermal chart, and evaluating the weak-deflection integral along the zeroth-order straight line
\begin{equation}
\gamma: \quad v=b,\qquad u\in[u_S,u_R],\qquad \rho=\sqrt{u^2+b^2},
\label{eq:straight_ray}
\end{equation}
we obtain
\begin{equation}
\alpha
=\int_{u_S}^{u_R}\hat n^{\,i}\partial_i\varphi\,du
= -\int_{u_S}^{u_R}\frac{b}{\rho}\,\varphi'(\rho)\,du.
\label{eq:alpha_integral_uv}
\end{equation}

The isothermal radius $\rho(r)$ is determined by equating the radial parts of \eqref{eq:Kottler_optical} and \eqref{eq:isothermal_conformal}, which yields
\begin{equation}
\frac{d\rho}{\rho}=\frac{dr}{r\sqrt{f(r)}}.
\label{eq:rho_ode}
\end{equation}
Expanding $1/\sqrt{f(r)}$ to first order in $M$ and $\Lambda$ gives
\begin{equation}
\frac{1}{\sqrt{f(r)}}=1+\frac{M}{r}+\frac{\Lambda}{6}r^2+\mathcal{O}(M^2,\Lambda^2,\Lambda M).
\label{eq:f_sqrt_expand}
\end{equation}
Integrating \eqref{eq:rho_ode} with \eqref{eq:f_sqrt_expand} yields
\begin{equation}
\ln\rho = \ln r -\frac{M}{r}+\frac{\Lambda}{12}r^2 + \text{const}
+\mathcal{O}(M^2,\Lambda^2,\Lambda M).
\label{eq:lnrho}
\end{equation}
Choosing the integration constant so that $\rho\sim r$ in the weak-field region, we obtain
\begin{equation}
\rho = r - M + \frac{\Lambda}{12}r^3 + \mathcal{O}(M^2,\Lambda^2,\Lambda M).
\label{eq:rho_of_r}
\end{equation}
Because Kottler spacetime is not asymptotically flat, the asymptotic Euclidean normalization of Sec. \ref{sec6.2} is not imposed here. Instead, the additive constant in \(\ln\rho\) is fixed by the local weak-field matching choice \(\rho\sim r\) in the finite lensing region, which is sufficient for the perturbative evaluation below; by Sec. \ref{sec6.1}, the physical deflection angle is insensitive to this residual normalization choice when all boundary terms are handled consistently.

For the total deflection through $\mathcal{O}(M)$, $\mathcal{O}(\Lambda)$, and $\mathcal{O}(\Lambda M)$ within the straight-ray evaluation scheme \eqref{eq:alpha_integral_uv}, it is sufficient to retain
\begin{equation}
\rho = r - M + \mathcal{O}(M^2,\Lambda).
\label{eq:rho_shift_only}
\end{equation}
The key point is that the $\mathcal{O}(\Lambda M)$ contribution to $\alpha$ will arise from substituting \eqref{eq:rho_shift_only} into the \emph{pure} $\mathcal{O}(\Lambda)$ endpoint term, as shown below.

Next, from \eqref{eq:Kottler_optical} and \eqref{eq:isothermal_conformal}, the angular coefficient gives
\begin{equation}
e^{2\varphi(\rho)}\,\rho^2=\frac{r(\rho)^2}{f(r(\rho))},
\qquad\Rightarrow\qquad
e^{2\varphi(\rho)}=\frac{r(\rho)^2}{f(r(\rho))\,\rho^2}.
\label{eq:e2varphi_exact}
\end{equation}
Expanding to first order in $M$ and $\Lambda$, and using $r(\rho)=\rho+\mathcal{O}(M,\Lambda\rho^3)$ at this step, we obtain
\begin{equation}
\varphi(\rho)=\frac{2M}{\rho}+\frac{\Lambda}{12}\rho^2+\mathcal{O}(M^2,\Lambda^2,\Lambda M).
\label{eq:varphi_expand}
\end{equation}
Differentiating gives
\begin{equation}
\varphi'(\rho)=-\frac{2M}{\rho^2}+\frac{\Lambda}{6}\rho+\mathcal{O}(M^2,\Lambda^2,\Lambda M).
\label{eq:varphi_prime}
\end{equation}

Next, we perform the evaluation of the ray integral. Substituting \eqref{eq:varphi_prime} into \eqref{eq:alpha_integral_uv} yields, to the stated order,
\begin{equation}
\alpha
=\int_{u_S}^{u_R}\left[
\frac{2Mb}{(u^2+b^2)^{3/2}}-\frac{\Lambda b}{6}
\right]\,du
+\mathcal{O}(M^2,\Lambda^2,\Lambda M).
\label{eq:alpha_split}
\end{equation}
The two elementary integrals are
\begin{equation}
\int \frac{du}{(u^2+b^2)^{3/2}}=\frac{u}{b^2\sqrt{u^2+b^2}},
\qquad
\int du = u.
\label{eq:integrals_used}
\end{equation}
Therefore
\begin{equation}
\alpha
=
\frac{2M}{b}\left[\frac{u}{\sqrt{u^2+b^2}}\right]_{u_S}^{u_R}
-\frac{\Lambda b}{6}\left[u\right]_{u_S}^{u_R}
+\mathcal{O}(M^2,\Lambda^2,\Lambda M).
\label{eq:alpha_uform}
\end{equation}

Writing $\rho_X\equiv\sqrt{u_X^2+b^2}$ for $X=S,R$, we have
\begin{equation}
\frac{u_X}{\rho_X}=\mathrm{sgn}(u_X)\sqrt{1-\frac{b^2}{\rho_X^2}},
\qquad
u_R-u_S=\sqrt{\rho_R^2-b^2}+\sqrt{\rho_S^2-b^2}.
\label{eq:u_rho_id}
\end{equation}
Using $u_R>0$ and $u_S<0$ gives
\begin{equation}
\alpha
=
\frac{2M}{b}\left[\sqrt{1-\frac{b^2}{\rho_R^2}}+\sqrt{1-\frac{b^2}{\rho_S^2}}\right]
-\frac{\Lambda b}{6}\left[\sqrt{\rho_R^2-b^2}+\sqrt{\rho_S^2-b^2}\right]
+\mathcal{O}(M^2,\Lambda^2,\Lambda M).
\label{eq:alpha_rho_endpoints}
\end{equation}

Finally, we will express the total angle in terms of the areal radii and recover the $\Lambda M$ term. To compare with the standard finite-distance expression, we re-express the endpoints in terms of the areal radii $r_S$ and $r_R$. To the order required, we use the universal shift \eqref{eq:rho_shift_only} at the endpoints:
\begin{equation}
\rho_X = r_X - M + \mathcal{O}(M^2,\Lambda),
\qquad X\in\{S,R\}.
\label{eq:rho_endpoint_shift}
\end{equation}
In the $\mathcal{O}(M)$ term of \eqref{eq:alpha_rho_endpoints}, replacing $\rho_X\to r_X$ changes $\alpha$ only at $\mathcal{O}(M^2)$, so we may write
\begin{equation}
\alpha_M
=
\frac{2M}{b}\left[\sqrt{1-\frac{b^2}{r_R^2}}+\sqrt{1-\frac{b^2}{r_S^2}}\right]
+\mathcal{O}(M^2,\Lambda M).
\label{eq:alpha_M_final}
\end{equation}
In the pure $\mathcal{O}(\Lambda)$ term, however, substituting \eqref{eq:rho_endpoint_shift} produces an additional $\mathcal{O}(\Lambda M)$ correction:
\begin{align}
\sqrt{\rho_X^2-b^2}
&=\sqrt{(r_X-M)^2-b^2}\nonumber\\
&=\sqrt{r_X^2-b^2}-\frac{M r_X}{\sqrt{r_X^2-b^2}}+\mathcal{O}(M^2).
\label{eq:sqrt_expand}
\end{align}
Hence
\begin{equation}
\alpha_\Lambda
=
-\frac{\Lambda b}{6}\left[\sqrt{r_R^2-b^2}+\sqrt{r_S^2-b^2}\right]
+\frac{\Lambda Mb}{6}\left[\frac{r_R}{\sqrt{r_R^2-b^2}}+\frac{r_S}{\sqrt{r_S^2-b^2}}\right]
+\mathcal{O}(\Lambda M^2,\Lambda^2).
\label{eq:alpha_Lambda_and_LambdaM}
\end{equation}
Combining \eqref{eq:alpha_M_final} and \eqref{eq:alpha_Lambda_and_LambdaM}, and writing $u_X\equiv 1/r_X$, yields the total deflection angle through $\mathcal{O}(M)$, $\mathcal{O}(\Lambda)$, and $\mathcal{O}(\Lambda M)$:
\begin{align}
\alpha
&=
\frac{2M}{b}\left[\sqrt{1-b^2u_R^2}+\sqrt{1-b^2u_S^2}\right]
-\frac{\Lambda b}{6}\left[\frac{\sqrt{1-b^2u_R^2}}{u_R}+\frac{\sqrt{1-b^2u_S^2}}{u_S}\right]\nonumber\\
&\quad\;
+\frac{\Lambda Mb}{6}\left[\frac{1}{\sqrt{1-b^2u_R^2}}+\frac{1}{\sqrt{1-b^2u_S^2}}\right]
+\mathcal{O}(M^2,\Lambda^2),
\label{eq:alpha_Kottler_total}
\end{align}
which reproduces the standard finite-distance Kottler expansion (including the explicit $\mathcal{O}(\Lambda)$ background contribution and the mixed $\mathcal{O}(\Lambda M)$ term) within the weak-deflection evaluation scheme used here \cite{Ishihara:2016vdc}.

As a final remark, Eq. \eqref{eq:alpha_boundary_only} is an exact consequence of the boundary-only isothermal reduction for the chosen domain and closure. The explicit expansion \eqref{eq:alpha_Kottler_total} further assumes the weak-field/weak-deflection evaluation along the straight reference ray \eqref{eq:straight_ray}. In particular, the mixed $\mathcal{O}(\Lambda M)$ term in \eqref{eq:alpha_Kottler_total} arises from consistently relating the isothermal radius $\rho$ to the areal radius $r$ at the endpoints in the presence of the $\mathcal{O}(\Lambda)$ contribution.

\section{Conclusion} \label{sec8}
We have developed a boundary only formulation of finite distance gravitational lensing within the Gauss-Bonnet approach by exploiting the special structure of two dimensional optical geometry in isothermal coordinates. Our central technical move is to place the equatorial optical metric into a conformal form and use the two dimensional identity that collapses the Gaussian curvature into a Laplacian acting on the conformal factor. This converts the curvature area contribution in the Gauss-Bonnet theorem into a pure boundary integral, as expressed in Eq. \eqref{4.22}, and yields a deflection formula in which the bulk geometry enters only through boundary data evaluated on the chosen lensing domain, as organized in Eq. \eqref{4.26}.

A key practical element is the finite distance domain construction. By choosing a domain bounded by the physical light ray, two radial segments anchored at the source and receiver, and a closing curve chosen naturally in the isothermal geometry, we can (i) maintain consistency with finite distance angle definitions of the Ishihara type and (ii) control or eliminate geodesic-curvature contributions. The isothermal-circle closure is particularly effective: it keeps the corner-angle bookkeeping transparent, preserves orthogonality properties under conformal rescaling, and enables cancellations that leave the deflection determined by a single boundary integral along the ray in the conformal plane in the static, axisymmetric cases. This mechanism was made explicit in the Schwarzschild example, where the closure and azimuthal bookkeeping cancel identically and the final leading-order deflection reduces to the ray integral in Eq. \eqref{7.16}.

The worked examples demonstrate that the formalism is not merely a change of variables but a genuine
computational simplification. In Schwarzschild, the reduction is exact at the geometric level: once
the optical metric is written in isothermal form, the leading finite distance bending follows from
elementary one-dimensional integrals along a straight reference ray, without computing $K$
explicitly and without performing any two dimensional area integral. The Reissner-Nordstr\"om
example shows that the same boundary only method cleanly separates the leading mass term from the
leading charge correction in a finite distance setting, again via explicit boundary evaluation
rather than curvature integration. Finally, the Kottler case illustrates that the boundary only
isothermal reduction remains applicable in a representative non-asymptotically flat geometry:
working with the operational finite-distance definition of the \emph{total} deflection angle,
$\alpha=\Psi_R-\Psi_S+\Phi_{RS}$, we recover the standard weak-field finite distance expansion
including the explicit $\mathcal{O}(\Lambda)$ term and the mixed $\mathcal{O}(\Lambda M)$ correction.
In non-asymptotically flat settings one may also introduce background-subtracted variants (e.g.\ by
subtracting the $M=0$ value at fixed $\Lambda$) if a lens-induced quantity is desired; such
subtractions, however, define observables distinct from the total angle and should be specified
separately.

Compared with curvature area computations, the present approach removes the most expensive step: evaluating \(\iint_D K\,dS\) over a domain whose boundary is itself determined by the ray geometry. In classical Gauss-Bonnet lensing this step typically requires an explicit curvature expression, an explicit area element, and a nontrivial integration over a ray-bounded region, often followed by delicate limiting procedures. Li-type finite distance computations already aim to reduce this workload by replacing the area integral with a one-dimensional integral after a radial primitive is constructed and normalized, commonly via a special radius such as a circular orbit. The boundary only isothermal method accomplishes the reduction structurally: the area term is converted to a boundary term by geometry alone (Eq. \eqref{4.22}), and normalization issues are handled by the intrinsic scaling freedom of the isothermal radius (Section \ref{sec6}) rather than by imposing orbit dependent conditions. Operationally, this shifts the analytic effort from \emph{integrate curvature over a moving domain} to \emph{evaluate conformal boundary data on simple curves}, which is precisely what makes the Schwarzschild and Reissner-Nordstr\"om derivations transparent.

The domain of validity of the formalism is the region where the equatorial optical geometry is Riemannian and where an isothermal chart can be constructed on the relevant lensing domain. Practically, this means working outside degeneracy surfaces of the optical metric (e.g., avoiding regions where the static optical construction fails) and choosing source/receiver configurations for which the domain can be taken simply connected with piecewise smooth boundary. In non-asymptotically flat settings, one must also specify the operational definition of \emph{deflection} relative to an appropriate reference geometry; the boundary only representation makes this dependence explicit and therefore easier to control.

Several research directions follow naturally from what we have established here. One direction is to extend the boundary only reduction beyond static metrics to genuinely stationary spacetimes in a systematic way, clarifying when and how an osculating Riemannian representative of the Randers optical geometry can be used so that the boundary only identities retain their utility while correctly capturing gravitomagnetic contributions. Another direction is to develop higher-order weak-field expansions within the same boundary framework in a fully controlled manner, including the consistent incorporation of first-order trajectory corrections that become necessary for second-order bending. A third direction is to explore more general background-subtraction and renormalization prescriptions in non-asymptotically flat optical geometries (including cosmological or plasma-modified settings), where \emph{lensing by the lens} must be defined relative to a physically meaningful reference propagation problem.

\acknowledgments
R. P. and A. \"O. would like to acknowledge networking support of the COST Action CA21106 - COSMIC WISPers in the Dark Universe: Theory, astrophysics and experiments (CosmicWISPers), the COST Action CA22113 - Fundamental challenges in theoretical physics (THEORY-CHALLENGES), the COST Action CA21136 - Addressing observational tensions in cosmology with systematics and fundamental physics (CosmoVerse), the COST Action CA23130 - Bridging high and low energies in search of quantum gravity (BridgeQG), and the COST Action CA23115 - Relativistic Quantum Information (RQI) funded by COST (European Cooperation in Science and Technology). A. \"O. also thanks to EMU, TUBITAK, ULAKBIM (Turkiye) and SCOAP3 (Switzerland) for their support.

\bibliography{ref}

\end{document}